\documentclass[useAMS,usenatbib,usegraphicx]{mn2e}
\usepackage{mathptmx}
\usepackage{amssymb}
\usepackage{color}
\bibpunct{(}{)}{;}{a}{}{,}

\newcommand{\ignore}[1]{}

\newcommand{\mghk}[1]{#1}
\newcommand{\com}[1]{\textcolor{green}{}}
\newcommand{\jmr}[1]{#1}
\newcommand{\pa}[1]{#1}
\newcommand{\change}[1]{#1}
\newcommand{\cmt}[1]{}

\newcommand{\rvone}[1]{{#1}}

\title[\rvone{Opening angle and morphology of extragalactic radio
sources}]{\rvone{A new connection between the opening angle and the \pa{large-scale}
morphology of extragalactic radio sources}}

\author[M.\ Krause et al.]{Martin Krause,$^{1,2,3}$\thanks{E-mail:
           {\tt krause@mpe.mpg.de}}
        Paul Alexander,$^{1,4}$,  Julia Riley$^1$ and Daniel Hopton$^1$\\
        $^1$Astrophysics Group, Cavendish Laboratory, 19~J.~J.~Thomson Avenue,
            Cambridge CB3 0HE\\
        $^2$Universit\"atssternwarte M\"unchen, Scheinerstr.~1,
            81679 M\"unchen, Germany\\
        $^3$Max-Planck-Institut f\"ur Extraterrestrische Physik,
            Giessenbachstrasse, 85748 Garching, Germany \\
        $^4$Kavli Institute for Cosmology Cambridge, Madingley Road,
        Cambridge, CB3 0HA}
\begin{document}

\date{Accepted .-.  Received .-.}

\pagerange{\pageref{firstpage}--\pageref{lastpage}} \pubyear{2009}

\maketitle

\label{firstpage}



\begin{abstract}
\jmr{In the case of an initially conical jet, we study the relation 
\change{between} jet
collimation by the external pressure and \pa{large-scale} morphology}. We first
consider the important \jmr{length-}scales \jmr{in} the problem, and then carry out
axisymmetric \jmr{hydrodynamic} simulations that include, for certain
parameters, all the\jmr{se length-}scales. We find three important
scales related to the collimation region: \jmr{(i)} where the
sideways ram-pressure equals the external pressure, \jmr{(ii)} where the
jet density equals the ambient density, and \jmr{(iii)} where the
forward ram-pressure falls below the ambient pressure. These scales
are set by the external Mach-number and opening angle \jmr{of the jet}. We
demonstrate that the relative magnitude\jmr{s} of these scales
determine \jmr{the}
collimation, Mach-number, density and morphology of the large
scale \mghk{jet. Based on analysis of the shock structure, we}
reproduce successfully the morphology of Fanaroff-Riley (FR) class~I and ~II
radio sources. \mghk{Within the framework of the model, an FR~I radio source
must have a large intrinsic opening angle. Entrainment of ambient gas
might also be important.}
We also show that all FR~I sources with radio lobes or similar
features must have had an earlier FR~II phase.
\end{abstract}


\section{Introduction}\label{sec:intro}
Extragalactic radio \jmr{sources} are traditionally divided into two
morphological classes \citep{FR74}: \jmr{those in} class~I \jmr{(FR~I)
  are} edge-darkened \jmr{and those in class~II} edge-brightened. 
\jmr{\citeauthor{FR74} found that t}he two classes differ in radio
power, \jmr{ t}hose \jmr{with} 178~MHz \jmr{luminosities below} $2\times
10^{25}$~W~Hz$^{-1}$~sr$^{-1}$ \jmr{(Hubble's constant = 50 km
  s$^{-1}$ Mpc$^{-1}$)} \jmr{being} in class~I \jmr{and those
  above} in class~II. In subsequent
studies, the critical power\jmr{, $P_\mathrm{crit}$,} was found to be correlated with the
optical magnitude of the host galaxy \citep[$P_\mathrm{crit}\propto
L_\mathrm{opt}^2$,][]{OL89,OW91,Owen93,LO96,GC01}.  
It is well known that brighter galaxies tend to have bigger
central super-massive black holes \citep[e.g.][]{Guea09}, which
are responsible for the jet
production in the first place. Interestingly \jmr{however},
\citeauthor*{WLA07} (\citeyear{WLA07}) do not find a correlation of
FR~class with the mass of the super-massive black hole in the host
galaxy for a sample of 21~3CRR radio sources near the critical 
power. Thus, \change{in determining} the large scale
morphology, the jet power is the most important 
\change{influence --} at least
one other also plays a role, but 
\change{this factor} does not seem to be the black hole mass.

FR~I radio sources often show narrow jets
emanating from the core, which is
usually thought to coincide with the AGN, and extending out to a bright flaring
point. Beyond this flaring point, the flow widens and dims. These
findings have been modeled as a transition from a laminar to a
turbulent flow \citep[\mghk{e.g.}][]{Bick84,Bick86a,Bick86b,Kom90b,Kom90a,Kom90c,Wangea09},
which mixes turbulently with the entrained
gas. 
This basic model has been confirmed by numerical simulations:
  \citet{Rossiea08} show that a jet that is forced to entrain may make the
desired transition. \citet{PM07} find the transition in flow character
after an induced strong expansion phase in the jet. The reason for such an expansion
phase and why it should occur at low jet power only, remains however
unclear.
\citeauthor*{Hardea05} (\citeyear{Hardea05}) and
\citeauthor{Jetea05} (\citeyear{Jetea05}) present studies of
morphological features in three FR~I sources, focusing on the density
structure in the ambient X-ray gas as observed by {\em Chandra} and
{\em XMM-Newton}. They find the
flaring point is located at a place where the ambient gas
temperature rises steeply from the inner cool core to the hotter
intra-cluster gas.  This is accompanied by a local flattening of the
density profile.  In each case, the location is a few tens of kpc from
the core. It is clear that observations suggest the FR~class of a
radio source is significantly influenced by both \change{the} intrinsic jet
properties and the properties of the environment.

\jmr{The physics of} the structure and \jmr{evolution} of FR~II
sources \jmr{has been studied in some detail}. Such sources often show 
evidence of a jet at various places from the core out to the edge 
of the source, which might be over a Megaparsec away (e.g. 
\citeauthor*{Mulea06} \citeyear{Mulea06}). There, they feature 
one or more hotspots. \jmr{A} diffuse,
cylindrical ``cocoon'' \jmr{extends} from the hot spots \jmr{back} towards the radio core. 
Axis ratios of these cocoons vary between one and about six 
(e.g. \citeauthor{KA97} \citeyear{KA97}, and references therein,
\citeauthor*{Mulea08} \citeyear{Mulea08}). This has inspired
self-similar models of FR~II radio source growth
\citep{Falle91,KA97}. These models are valid far away from the 
characteristic scales of the problem. An outer scale $L_2$ \citep{KF98}, \jmr{at
  around which the source comes into pressure equilibrium with its
surroundings,} is given by
\begin{eqnarray}
L_2 &=& \left(\frac{Q_0}{\rho_\mathrm{x}\,c_\mathrm{x}^3} \right)^{1/2}\\
 &=& 324 \,\mathrm{kpc}\,  \nonumber
\left(\frac{Q_0}{10^{39}\,\mathrm{W}}\right)^{1/2}
\left(\frac{\rho_\mathrm{x}}{10^{-23}\,\mathrm{kg}\,\mathrm{m}^{-3}}\right)
^{-1/2}\\
 &\,& \times \nonumber
\left(\frac{c_\mathrm{x}}{1000\,\mathrm{km\,s^{-1}}}\right)^{-3/2} \, ,
\end{eqnarray}
where $Q_0$ is the kinetic power of the source, $\rho_\mathrm{x}$ is the 
external density, and $c_\mathrm{x}$ is the external sound speed. 
\mghk{FR~II r}adio sources approaching $L_2$ have been studied 
analytically \cmt{[Paul wanted Alexander 2006 as reference, but I
  think there must be some misunderstanding and so I leave Alexander
  2002]}
\citep{Alex02} and numerically 
(e.g. \citeauthor{Krause2005b} \citeyear{Krause2005b};
\citeauthor*{Gaiblea09} \citeyear{Gaiblea09}).
Up to $L_2$, the source is overpressured with respect to its environment,
and \change{is} also expected to have a strong bow shock (weakening as the source 
approaches $L_2$).
Observations of weak bow shocks in radio
sources $\approx 100$~kpc in size \citep[e.g.][]{McNamea05,Nulsea05}
may be simply understood in terms of them being close to $L_2$. 

Crucial to the existence of the self-similar
solution is the self-confinement of the jet by the source's own
cocoon pressure. 
This happens automatically in numerical models
of the propagation of light jets (heavy jets show only \jmr{a rudimentary}
cocoon).
In such simulations, the overpressured cocoon drives a re-collimation
shock into the beam as soon as the latter enters the computational
domain, \jmr{regardless of whether the beam is initially conical or
  cylindrical \mghk{\citep[e.g.][]{KF98,mypap03a,HKA07}}.}

An initially overdense, conical beam,
may be collimated via a re-collimation shock.
It occurs when the external pressure $p_\mathrm{x}$
becomes comparable to the sideways ram-pressure,
$\rho_\mathrm{j} v_\mathrm{j}^2 \sin^2\theta$, where $\rho_\mathrm{j}$ is the
jet density ($\propto r^{-2}$ in a conical jet, where r is the distance 
to the AGN), $v_\mathrm{j}$ the constant beam velocity, and
$\theta$ the half
opening angle (\citeauthor{Scheuer74}'s (\citeyear{Scheuer74}) model~B). 
This position is related to and occurs somewhat downstream
of the inner scale $L_1$ \citep[e.g.][]{Alex06}, given by
\begin{eqnarray}
L_1 &=& 2 \sqrt{2}\left(\frac{Q_0}{\rho_\mathrm{x}\,v_\mathrm{j}^3} \right)^{1/2}\\
\frac{L_1}{\mathrm{pc}} &=& 56 \, \nonumber
\left(\frac{Q_0}{10^{39}\,\mathrm{W}}\right)^{1/2}
\left(\frac{\rho_\mathrm{x}}{10^{-22}\,\mathrm{kg}\,\mathrm{m}^{-3}}\right)^{-1/2}
\left(\frac{v_\mathrm{j}}{c}\right)^{-3/2} \, .
\end{eqnarray}
On scales $L_1\ll L \ll L_2$, we expect self-similar behaviour.
At $L_1$ the jet density becomes roughly comparable to the external density, and
the mass of the swept-up gas becomes comparable to the mass that has gone 
through the jet channel.

\jmr{The development of the} large-scale morphology of the radio
source depends upon
what happens around $L_1$. If a proper 
re-collimation shock \jmr{is driven} into the beam, an FR~II source
\jmr{will develop}. \mghk{Otherwise, a morphology of the FR~I type might result.}

Here, we present jet simulations on a spherical grid starting at
a fraction of $L_1$ out to  a few hundred times \mghk{$L_1$}.
\jmr{We follow the approach of \citep{Alex06} in employing non-relativistic
hydrodynamics, addressing the same scale, downstream of
$L_1$.} This allows detailed
studies of the re-collimation process and its relation to the large-scale 
morphology. We show that
a proper re-collimation shock \jmr{only} occurs for small opening angles.

We further discuss the important scales of the problem in
Section~\ref{sec:scales}. Numerics and simulation setups are presented in 
Section~\ref{sec:sims}. The simulation results are shown in
Section~\ref{sec:res}, and discussed in Section~\ref{sec:disc}. We conclude
in Section~\ref{sec:conc}.

\section{Re-collimation, cocoon formation and terminal shock
  limit}\label{sec:scales}
\cmt{This section is based on Paul's comments to an earlier version of
the manuscript.}
\begin{table*}
\centering
    \begin{minipage}{130mm}
   \caption{Definition of length-scales \label{tab:scales}}
   \begin{tabular}{@{}llll@{}}

     \hline\hline
     Length-scale & formula & symbol & 
     assoc. transition\footnote{Symbol of the associated transition point when
     the jet is in a cocoon.}\\
     \hline
     
     Inner  & 
     $\left(\frac{8Q_0}{\rho_\mathrm{x}\,v_\mathrm{j}^3}\right)^{1/2} $&
     $L_1$ \\
     Re-collimation & 
     $\gamma^{1/2} M_\mathrm{x}\sin{\theta} L_1/(2 \Omega^{1/2})$&
     $L_\mathrm{1a}$ & $x_\mathrm{1a}$\\
     Cocoon formation &
     $ L_1/(2\Omega^{1/2})$&
       $L_\mathrm{1b}$ & $x_\mathrm{1b}$\\
     Terminal shock &
     $\gamma^{1/2} M_\mathrm{x} L_1/(2\Omega^{1/2})$&
       $L_\mathrm{1c}$ & $x_\mathrm{1c}$\\
       Outer  &
       $\left(\frac{Q_0}{\rho_\mathrm{x}\,c_\mathrm{x}^3} \right)^{1/2}$&
        $L_\mathrm{2}$\\
    
  \hline\hline
   \end{tabular}
   \end{minipage}
\end{table*}
 The physics of the region downstream of $L_1$ involves understanding
of the three length-scales 
(see \citet{Alex06} for further discussion, Table~\ref{tab:scales}
summarises the definitions),
associated to re-collimation
($L_\mathrm{1a}$), cocoon formation ($L_\mathrm{1b}$) and the terminal
shock ($L_\mathrm{1c}$). 
The first, $L_\mathrm{1a}$, is where the jet's sideways
ram-pressure equals the ambient pressure leading to potential
re-collimation by the ambient gas:
\begin{equation}
\left(\frac{L_\mathrm{1a}}{L_\mathrm{1}}\right)^2=
 \frac{\gamma}{4 \Omega} \,M_\mathrm{x}^2\sin^2\theta\, ,
\end{equation}
where $\gamma$ is the ratio of specific heats in the environment,
$\Omega=2 \pi (1-\cos\theta)$, the solid angle of the jet, and
$M_\mathrm{x}$ the Mach-number with respect to the ambient sound
speed. If, however, an overpressured cocoon forms, the jet-collimation
can occur on smaller scales. Before the jet collimates, the jet
density decreases with distance due to its fixed opening angle.
The jet starts to develop a cocoon at the second length-scale
$L_\mathrm{1b}$ where the density of the jet falls below that
of its environment:
\begin{equation}
\left(\frac{L_\mathrm{1b}}{L_\mathrm{1}}\right)^2=
 \frac{1}{4 \Omega} .
\end{equation}
The third, $L_\mathrm{1c}$ is the length-scale where the ram-pressure of the
un-collimated conical jet would fall \rvone{below the pressure in the
environment 
(see below for the case when an overpressured cocoon is present):} 
\begin{equation}
\left(\frac{L_\mathrm{1c}}{L_\mathrm{1}}\right)^2=
 \frac{\gamma}{4 \Omega} \,M_\mathrm{x}^2\, .
\end{equation}
Since the terminal shock is advanced by the \change{jet's} forward ram-pressure,
it will advance only up to $L_\mathrm{1c}$, unless the jet is
collimated before \rvone{reaching this third scale. If an
  overpressured cocoon is present, the critical scale for the jet
  termination shock will be further in. However, the scales for
  re-collimation and jet termination will be affected in the same way
  (compare eqs (3) and (5), the dependence on the environmental
  pressure is via the sound speed which enters the Mach number and for
  this consideration the environment would be the cocoon).} 
The terminal shock is the
prime site of particle acceleration and therefore synchrotron emission.
It may therefore be
identified with the hotspots of a radio source. If it remains at
$L_\mathrm{1c}$ because the jet fails to collimate beforehand, the
cocoon may still expand and the stationary shock at $L_\mathrm{1c}$ may
be identified with the flaring point of an FR~I source. 

It is interesting to look at the relation of these scales to one
another. The ratio $L_\mathrm{1b}/L_\mathrm{1a}$ is given by:
\change{
\begin{equation}
L_\mathrm{1b}/L_\mathrm{1a} = \frac{1}{\sqrt{\gamma} \sin \theta
M_\mathrm{x}}\, .
\end{equation}}
If $L_\mathrm{1a}<L_\mathrm{1b}$ then the jet will be collimated by
the external pressure before the jet density has fallen to that of the
surroundings. Once collimated the jet density does not decline further
along its length. Therefore, for $L_\mathrm{1a}<L_\mathrm{1b}$ a
cocoon does not form. Conversely, for $L_\mathrm{1a}>L_\mathrm{1b}$ a
cocoon is formed first and the jet is collimated by the pressure in
the cocoon. This second case is studied in detail by
\citet{KF98} and \citet{Alex06},
and the source transitions to near self-similar subsequent evolution.
The ratio
\begin{equation} \label{eta}
\eta=\left(\frac{L_\mathrm{1b}}{L_\mathrm{1a}}\right)^2 =
\frac{1}{\gamma \sin^2 \theta M_\mathrm{x}^2}\, .
\end{equation}
gives the ratio of the collimated jet density to the external
density. For a cocoon to form, $\eta<1$, and for a substantial cocoon,
$\eta<<1$. From Very Long Baseline Interferometry (VLBI), it is clear
that the jets are initially relativistic \citep[e.g.][]{Britzea08}\change{;
on scales of $\approx 100$~kpc, their velocities are still}
a reasonable fraction of the speed of light
\citep[e.g.][]{Hardea99,MH09}. Hence,
the external Mach-number for a jet is typically of order several
hundred. For very small opening angles of $\approx 1$~degree,
therefore a cocoon will not form. However, for larger opening angles
cocoons will form.

The ratio $L_\mathrm{1c}/L_\mathrm{1b}$ \change{is given by}:
\begin{equation}\label{L1cb}
L_\mathrm{1c}/L_\mathrm{1b} = \sqrt{\gamma} M_\mathrm{x}\, .
\end{equation}
\change{If $L_\mathrm{1c}/L_\mathrm{1b} >> 1 $ then a lobed
source, i.e. one that has formed a cocoon, also has the hotspots near
the leading edges.}
From Equation~(\ref{L1cb}), this is always the case. Hence, every radio source that
has lobes went through a phase when it had its hotspots
near the leading edges, i.e. it was once an FR~II source.
To remain an FR~II source, the re-collimation shock needs to reach the
axis, before the \mghk{jet} runs out of thrust. Calculation of this point is
difficult.
\citet{Alex06} argues that \rvone{re-collimation, in the case when the
  radio source is collimated by the pressure of a cocoon, should be
  completed at two to three times a length scale $L_\mathrm{r}$. 
$L_\mathrm{r}$ corresponds
  to $L_\mathrm{1a}$ in the present paper, and the numerical factor is in
  excellent agreement with the results of the simulations we present
  below.}
Adopting a value of 2.5, the critical ratio becomes:
\begin{equation}\label{L1ca}
L_\mathrm{1c}/2.5L_\mathrm{1a} = 1/(2.5\sin\theta)\, .
\end{equation}
Thus, jets with a half opening angle below about $24^\circ$ have enough
thrust to propel the terminal shock past the re-collimation shock.
Once this occurs, the terminal shock may
continue essentially indefinitely, since the beam is collimated,
and the ram-pressure stays constant (the exception being an inverted
pressure profile in the external medium). Such sources will end up as
large-scale FR~II radio sources. 

Jets with a half opening angle greater than $24^\circ$ have the 
re-collimation shock so far downstream that the terminal shock is
unable to reach it. In this case, the hotspot will stall at $L_\mathrm{1c}$,
and might turn into a flaring point. 
The exhaust plasma that leaves the beam may
still form a classical cocoon for some time. The plasma downstream of
the flaring point streams transsonically towards the tip of the
cocoon. There is no clear distinction between jet and cocoon in this region.
Ram-pressure balance between this plasma and the shocked
  ambient gas gives the velocity \change{$v_\mathrm{cd}$} of the
  contact surface \change{as}
  $v_\mathrm{cd}=\sqrt{\eta^\prime} v_\mathrm{coc}$ where
  $\eta^\prime$ is the density ratio between the cocoon and the
  shocked ambient gas, and $v_\mathrm{coc}$ is the bulk velocity of
  the cocoon plasma. Therefore, as long as the cocoon plasma is
  underdense with respect to the shocked ambient gas, the contact
  surface must advance more slowly than the cocoon plasma. The cocoon
  plasma therefore build\change{s} up thermal pressure near its leading edge and
  form\change{s} a backflow in a part of the cocoon. The situation changes at
  some point because of entrainment of shocked ambient gas into the
  cocoon. When enough dense gas is mixed into the cocoon so that
  -- spatially averaged --
  $\eta^\prime \approx1$, the contact surface advances at the same velocity
  as the cocoon plasma. The morphology is then transformed into a
  chimney-like outflowing structure. 
 
\begin{table*}
\centering
    \begin{minipage}{\textwidth}
   \caption{Simulation parameters \label{tab:simpars}}
   \begin{tabular}{@{}lrrrrrrrrrrrr@{}}

   \hline\hline
   Run      
   & $M_x$\footnote{External Mach-number} 	
   & $\theta$\footnote{Half opening angle / degree}  
   & $R_\mathrm{in}$\footnote{Inner boundary of the
     computational domain} / $L_1$
   & $R_\mathrm{f}$\footnote{Final size} / $L_1$
   & $L_\mathrm{1a}$\footnote{Length-scale where the sideways ram
     pressure equals the ambient pressure (re-collimation)}$/L_1$
   & $x_\mathrm{1a}$\footnote{Length-scale where the sideways ram
     pressure equals the cocoon pressure at the end of the simulations}$/L_1$
   & $L_\mathrm{1b}$\footnote{Length-scale where the density of
     the uncollimated jet would equal the ambient density (cocoon formation}$/L_1$
   & $L_\mathrm{1c}$\footnote{Length-scale where the forward ram
     pressure of the uncollimated jet would equal the ambient pressure
   (terminal shock limit)}$/L_1$
   & $x_\mathrm{1c}$\footnote{Length-scale where the forward ram
     pressure of the uncollimated jet would equal the cocoon pressure
     at the end of the simulations}$/L_1$
   & $L_\mathrm{2}$\footnote{Outer scale}$/L_1$
   & $dr$\footnote{Radial resolution (max.)}$/L_1$
   & $d\theta$\footnote{Meridional resolution (max.) / degree}\\
   \hline

  M05 & 500 &   5 & 0.363 &210 & 182 & 11 & 3.23 & 2087 &130 & 3953 & 0.142 & 0.088 \\ 
  M15 & 500 & 15 & 0.125 &  76 & 181 & 11 & 1.08 &   697 &  41 & 3953 & 0.049 & 0.352 \\ 
  M30 & 500 & 30 & 0.062 &  50 & 176 &   9 & 0.54 &   352 &  18 & 3953 & 0.024 & 0.352 \\ 
  m05 &     5 &   5 & 0.262 &  22 & 1.82 &    & 3.23 & 20.87 &      & 3.95 & 0.030  & 0.044 \\ 
  m15 &     5 & 15 & 0.098 &  18 & 1.81 &    & 1.08 &   6.98 &      & 3.95 & 0.039 & 0.022 \\ 
  m30 &     5 & 30 & 0.050 &  16 & 1.76 &    & 0.54 &   3.52 &      & 3.95 & 0.019 & 0.059 \\ 

   \hline           
  \hline
   \end{tabular}
   \end{minipage}
\end{table*}

\begin{figure*}
  \centering
       \includegraphics[width=.48\textwidth]{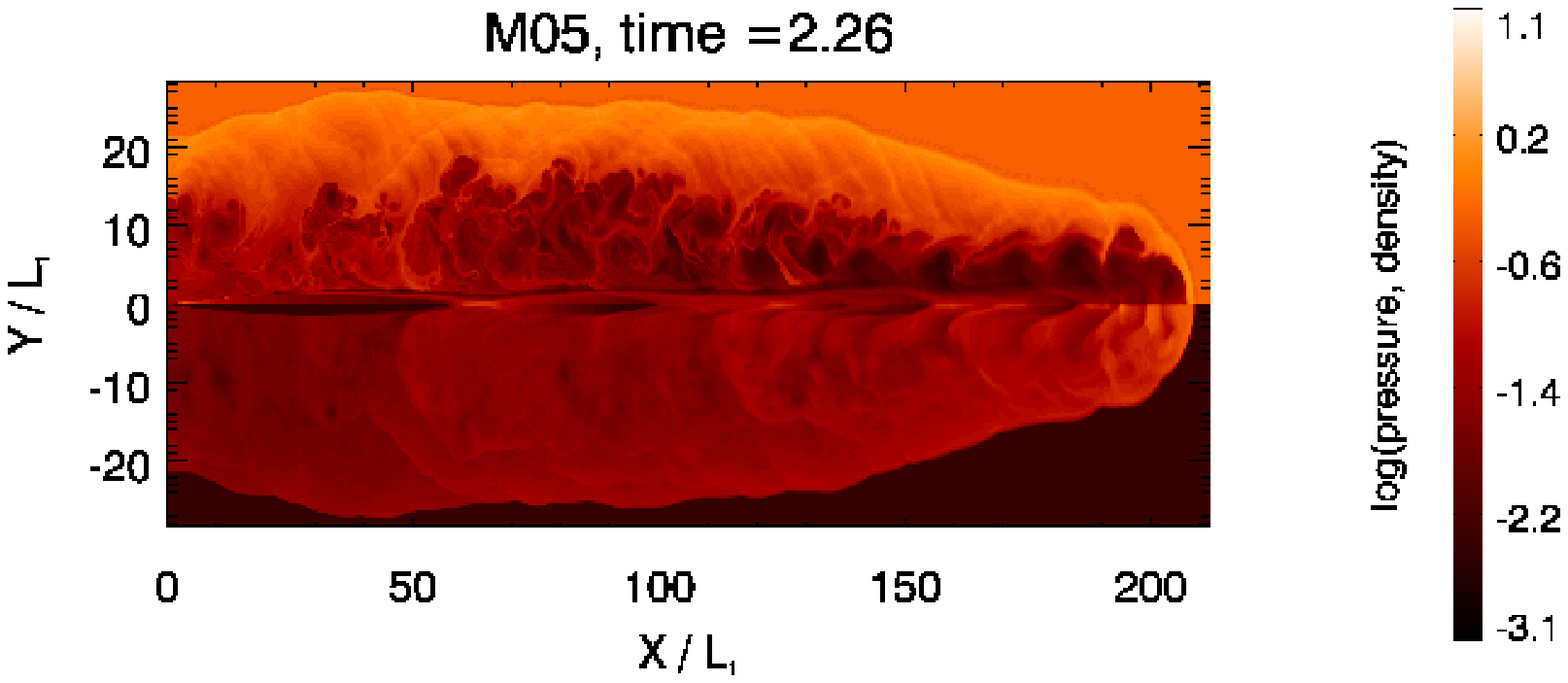} 
       \includegraphics[width=.48\textwidth]{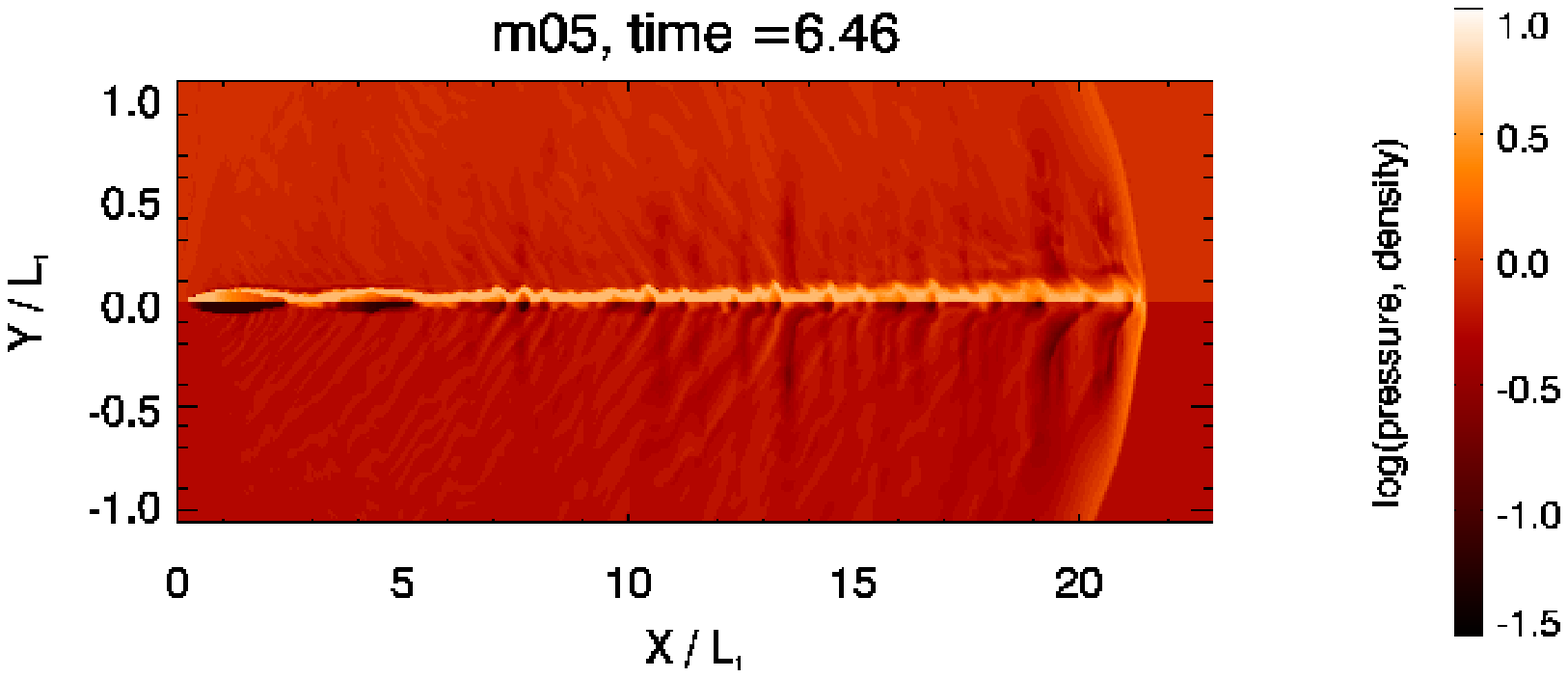} \\
       \includegraphics[width=.48\textwidth]{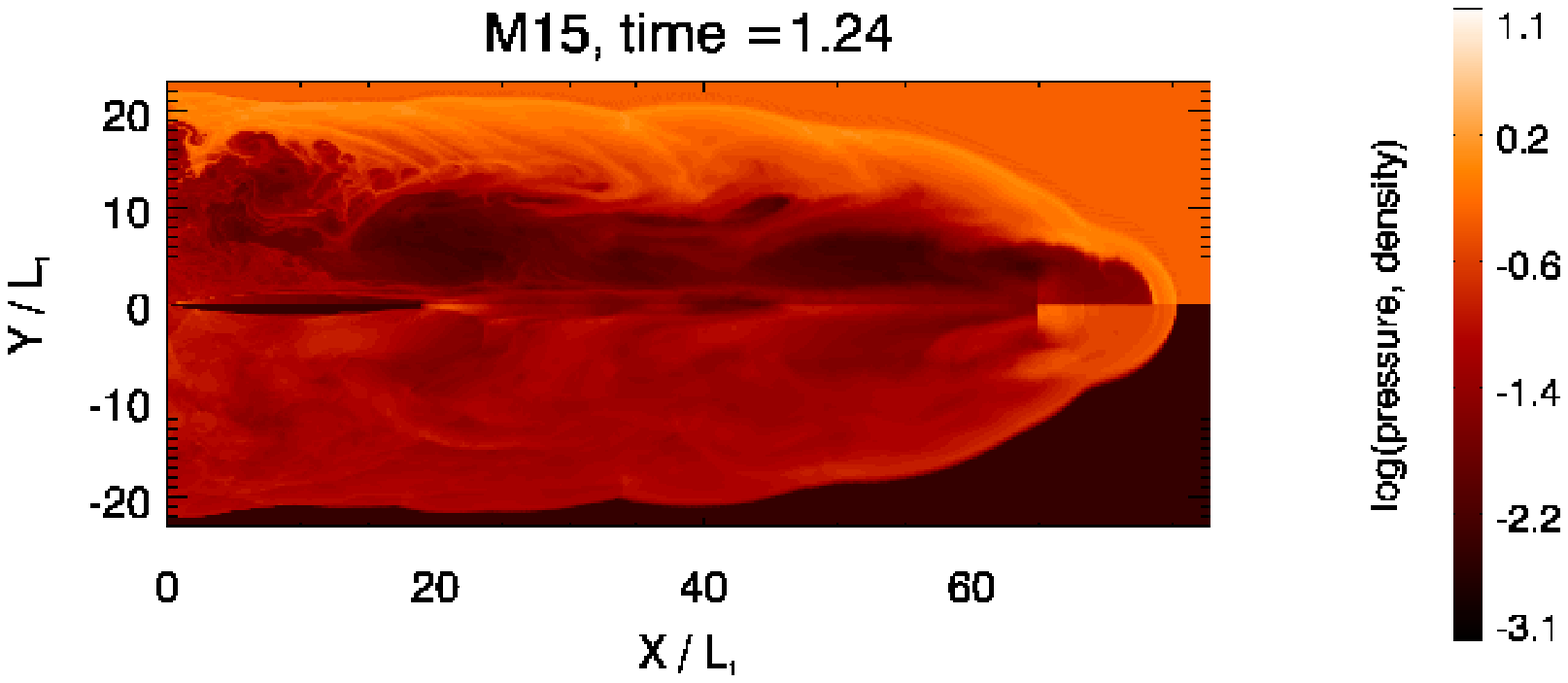} 
       \includegraphics[width=.48\textwidth]{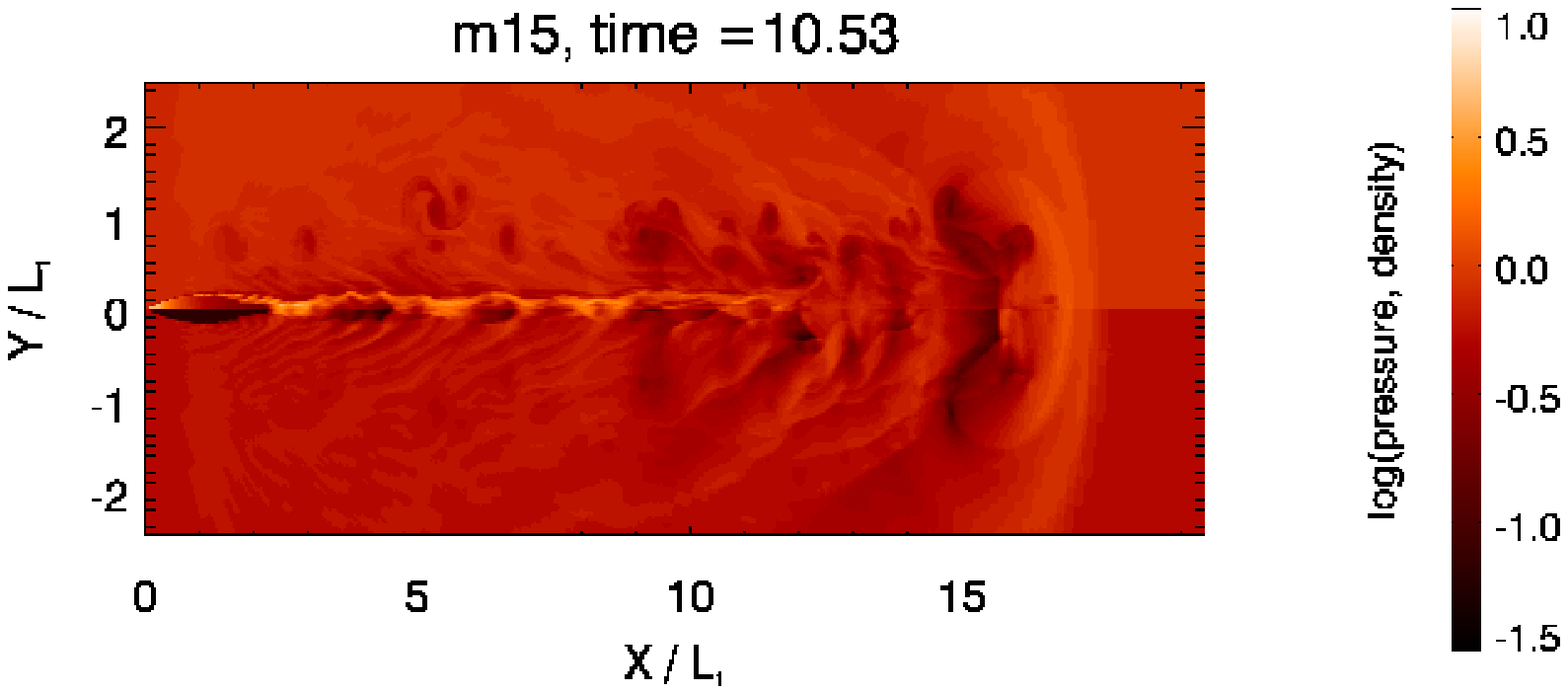} \\ 
       \includegraphics[width=.48\textwidth]{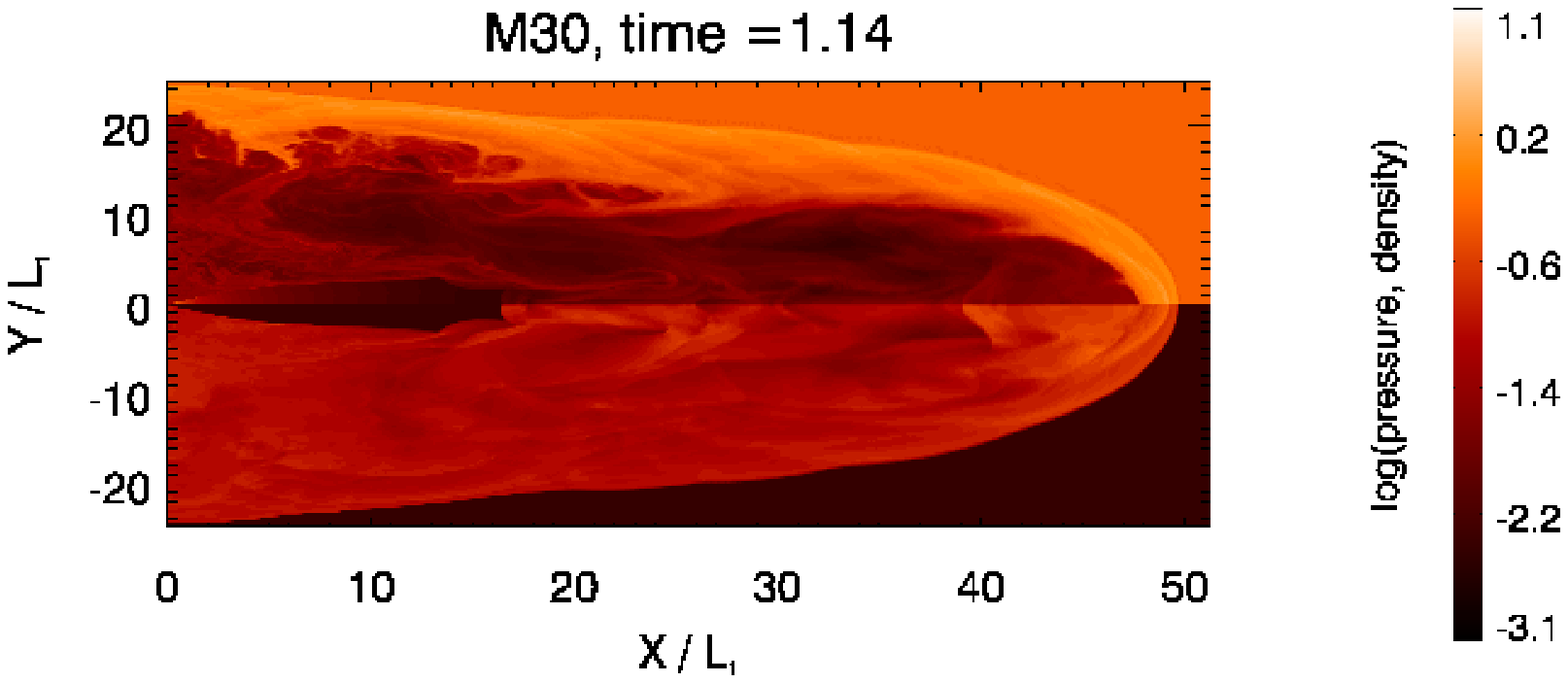} 
       \includegraphics[width=.48\textwidth]{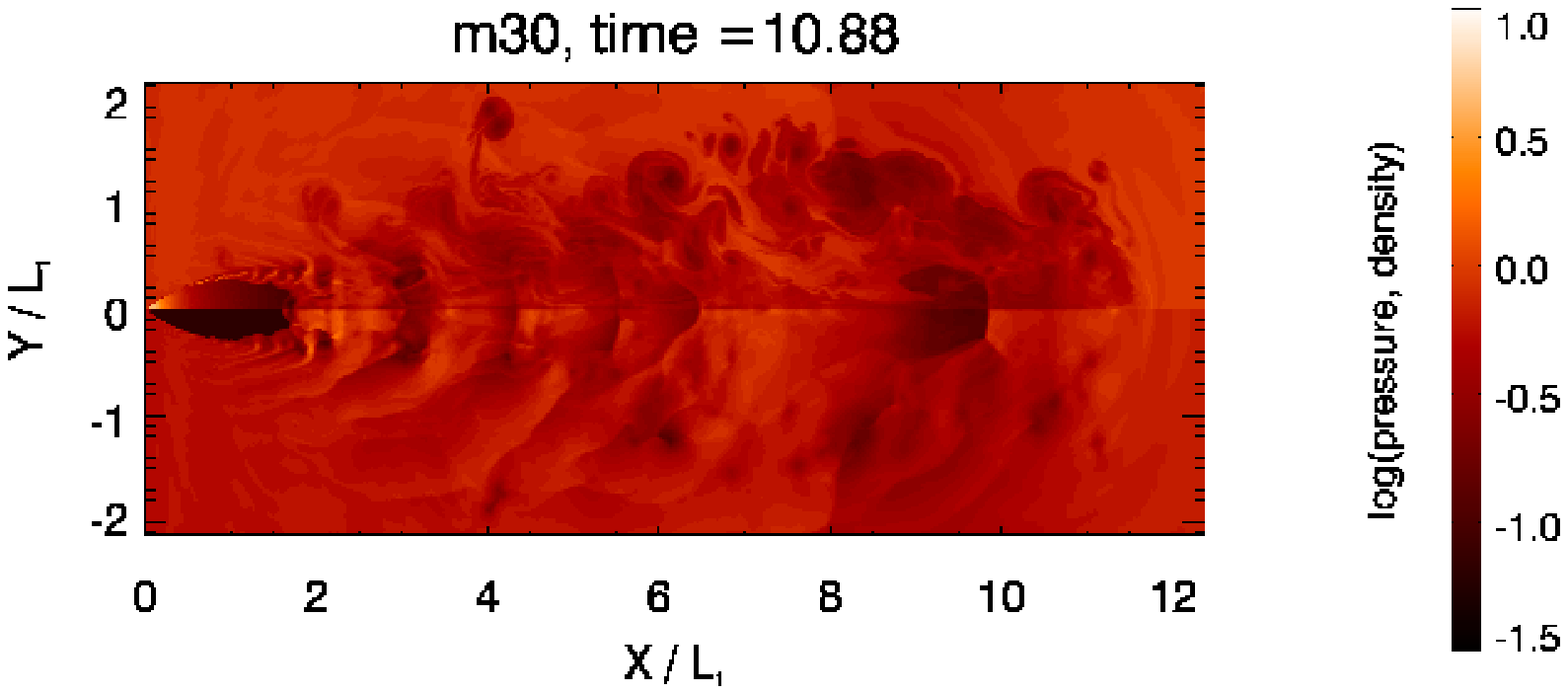} \\
     \vspace*{0pt}
 \caption{ Final logarithmic density \change{(top half) and pressure
     (bottom half)} distributions for all simulations. The
   time unit is $L_1$ divided by the ambient sound speed. The
   plots use different $x$ and $y$ scales in order to make the beam features visible. For
   run~m30, the density scale \mghk{saturates} at ten to enhance
   contrast, which affects the beam
   near the injection region, only.}
\label{fig:lgd}
\end{figure*}

\section{Numerics}\label{sec:sims}

We have carried out 2.5D axisymmetric hydrodynamics simulations with
the {\sc FLASH} code \citep[v. 2.5][]{Fryxea00,Caldea02} to
demonstrate the physics outlined in the previous section. {\sc FLASH}
evolves the equations of compressible hydrodynamics conservatively,
using third order (piecewise parabolic) interpolations. We have run
six simulations of conical jets into a constant ambient
density,  chosing units
of $L_1$ (length), $\rho_\mathrm{x}$ (external density),  and
$c_\mathrm{x}$ (external sound speed). The ratio of specific heats is
$\gamma=5/3$. In these units, the external pressure is therefore $3/5$.

We use spherical polar coordinates and the jet is implemented
as a boundary condition. Other boundaries are reflective.
In Section~\ref{sec:scales}, we found that the opening angle and the
Mach-number determine the characteristics of the flow. 
Therefore, we vary these parameters in the
simulations. The other parameter that is required to define the
  simulation setup is
the pressure ratio \mghk{between the jet and the ambient medium}. 
We set it to unity \mghk{at the jet injection boundary}, noting that the adiabatic
expansion renders the beam pressure quickly insignificant.
Finally, the computational domain specifies the outcome of the
simulation. Our aim is to include all the relevant scales of
Section~\ref{sec:scales}. However, choosing a realistic external Mach
number of~500\mghk{, and starting the jet smaller than $L_1$, we were unable to run
the simulation for long enough for the jet to reach $L_2$}, as three to four
orders of magnitude between the smallest and largest scales are
required for the three opening angles we chose. Making use of the
polar coordinates and the adaptive mesh refinement, \mghk{we reach a few
hundred times $L_1$}. Therefore, we have chosen a second set of
simulations with an external Mach-number of~5. For this second set,
all the relevant scales, including $L_2$, are within the computational
domain\mghk{, and we reach four to five times $L_2$. We choose three
  opening angles around the critical one defined in Equation~(\ref{L1ca}). } 
We label the simulations with a capital (small)~M for the high
and low Mach-number cases, respectively, followed by the half opening
angle in degrees. The simulation parameters are summarised in
Table~\ref{tab:simpars}.

\begin{figure*}
  \centering
  \includegraphics[width=\textwidth]{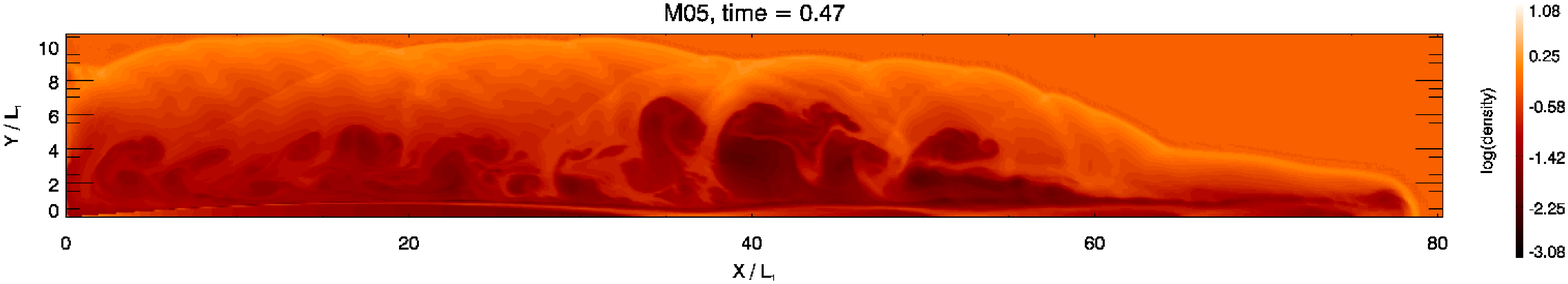} \\
  \includegraphics[width=\textwidth]{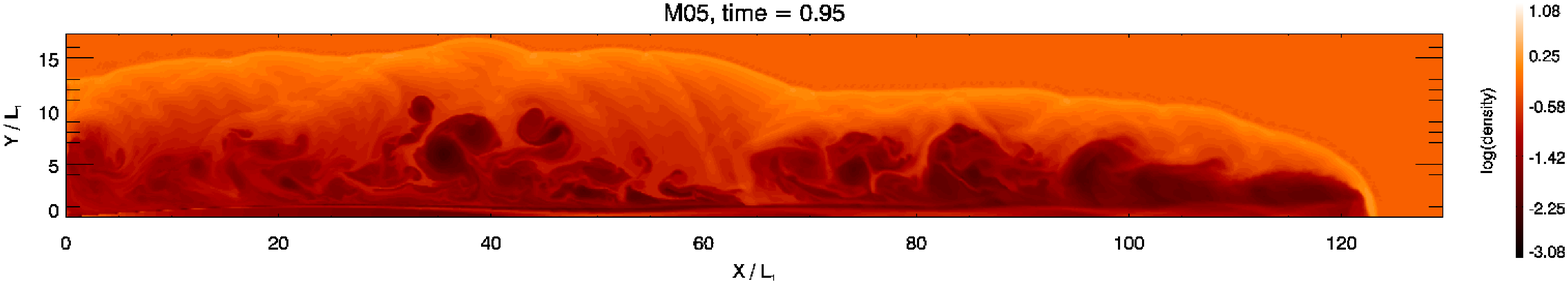} \\
  \includegraphics[width=\textwidth]{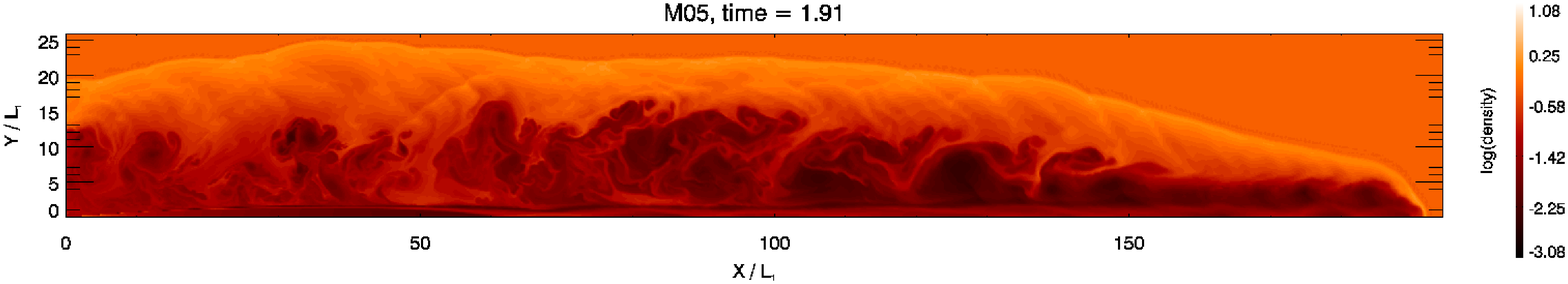} \\
 \vspace*{0pt}
 \caption{Logarithmic density distribution for run M05 for three snapshot times
   indicated in the title of the individual images. This jet always
   shows a clear FR~II morphology.}
\label{fig:M05evol}
\end{figure*}
\begin{figure*}
  \centering
  \includegraphics[width=.48\textwidth]{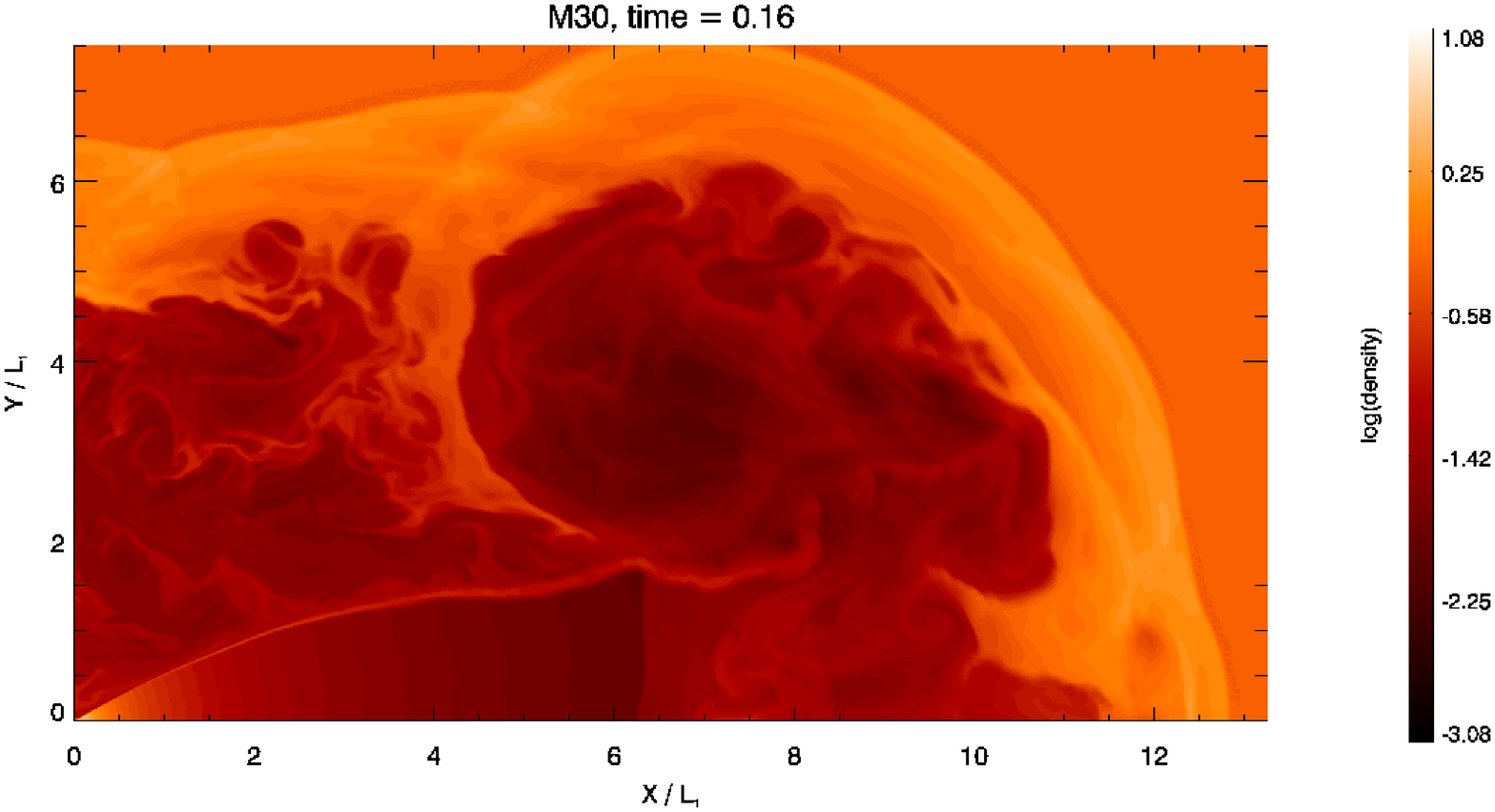} 
  \includegraphics[width=.48\textwidth]{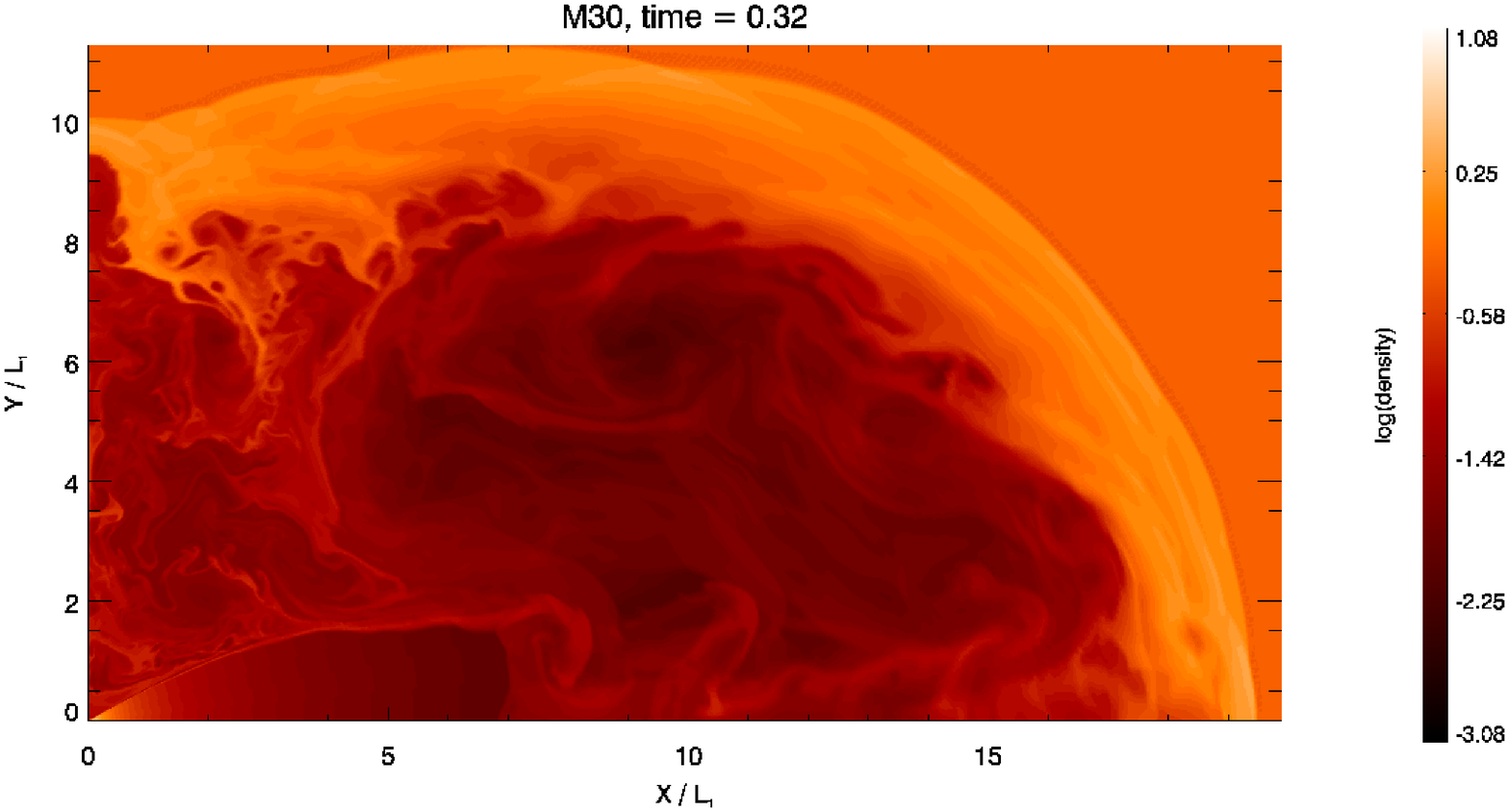} \\
  \includegraphics[width=.48\textwidth]{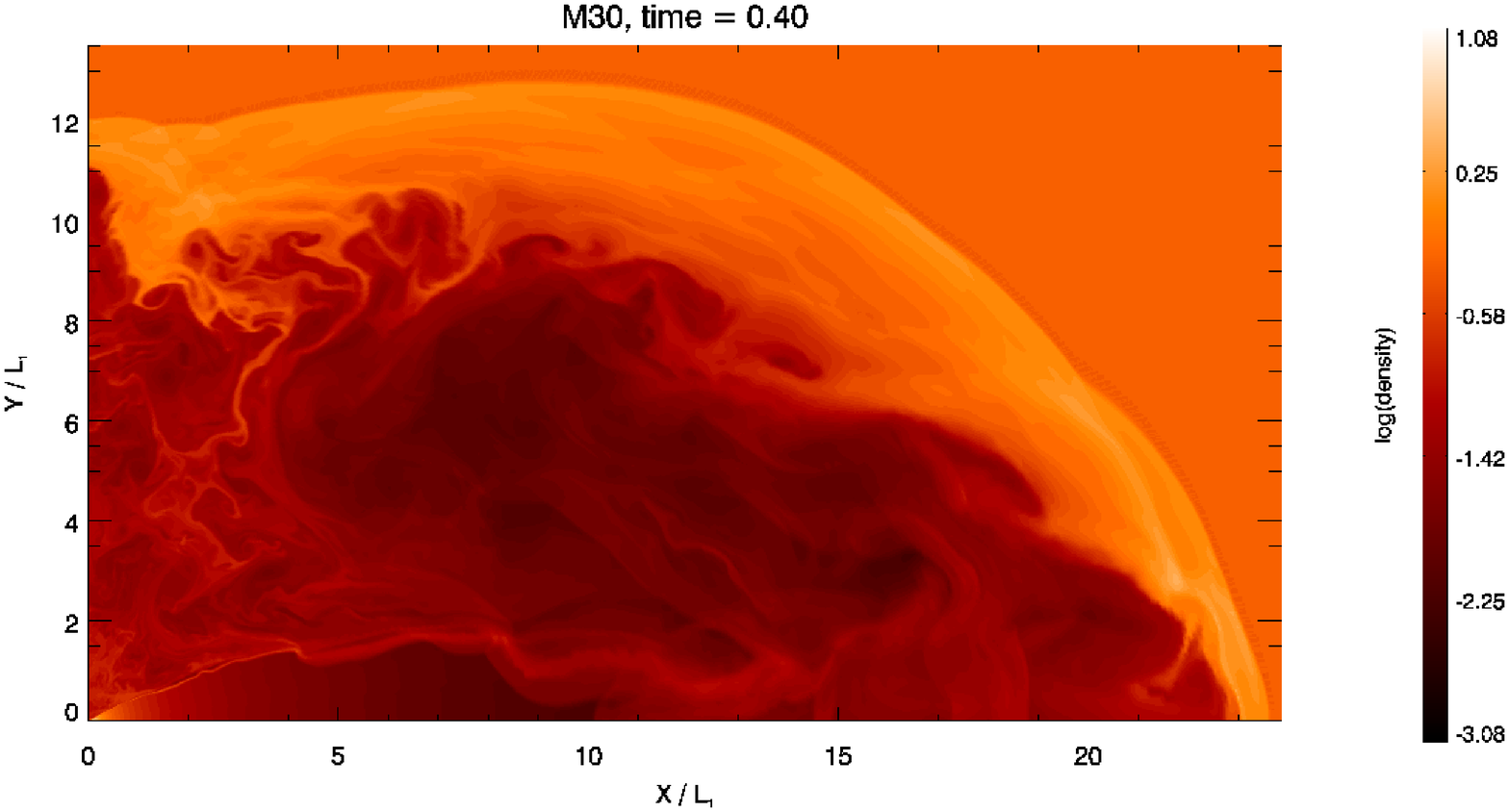} 
  \includegraphics[width=.48\textwidth]{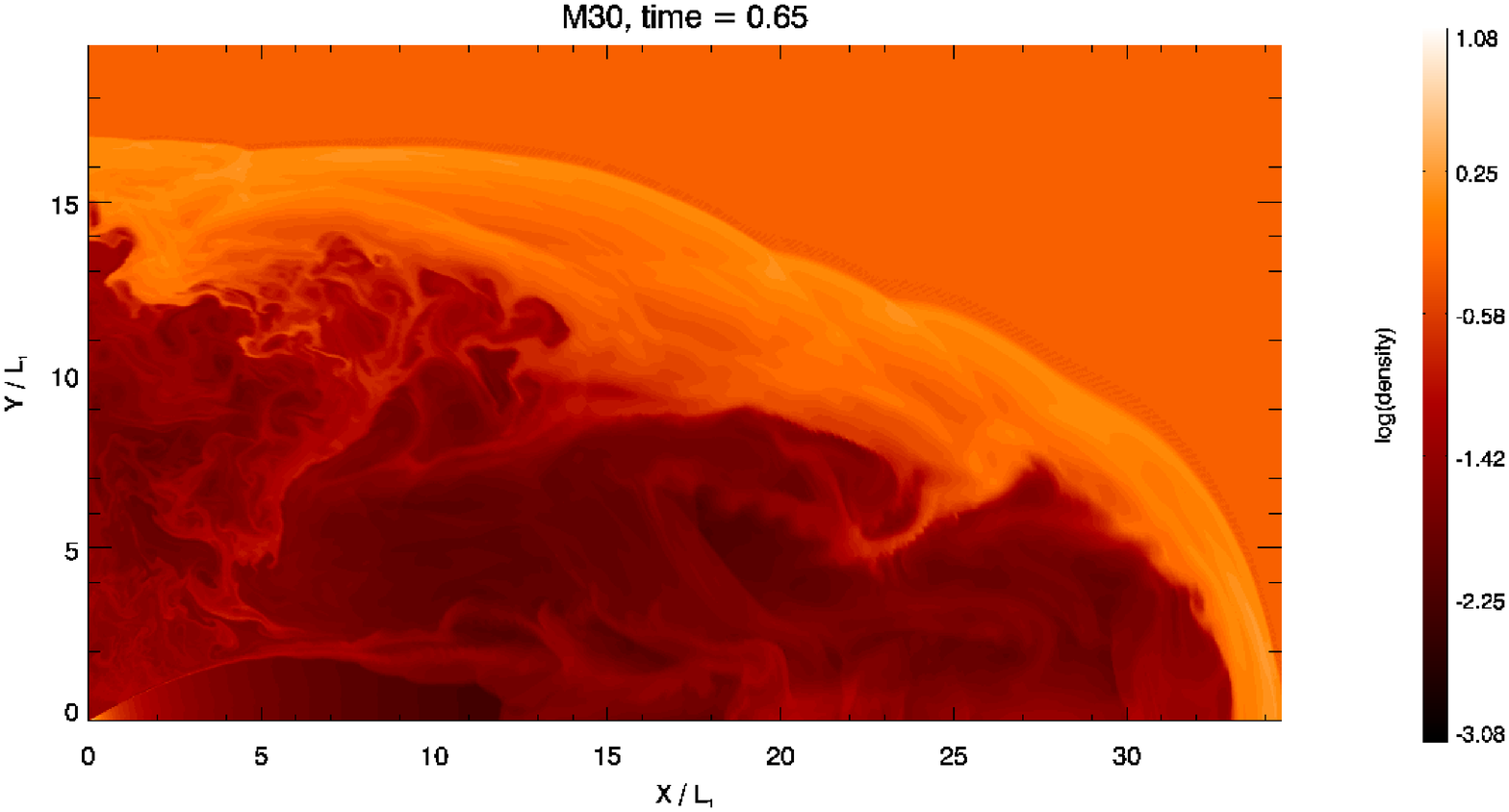} \\
  \includegraphics[width=.48\textwidth]{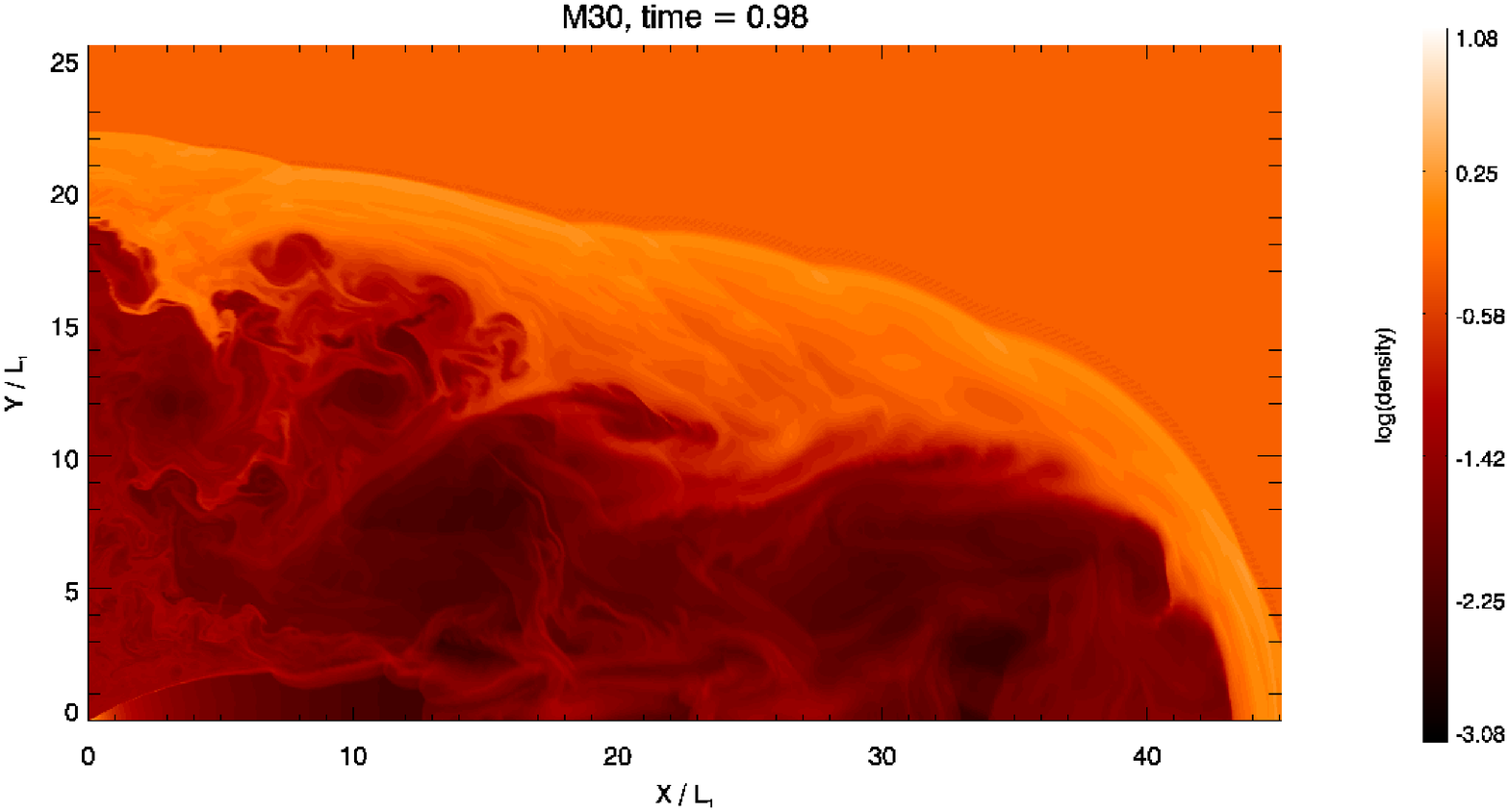} 
  \includegraphics[width=.48\textwidth]{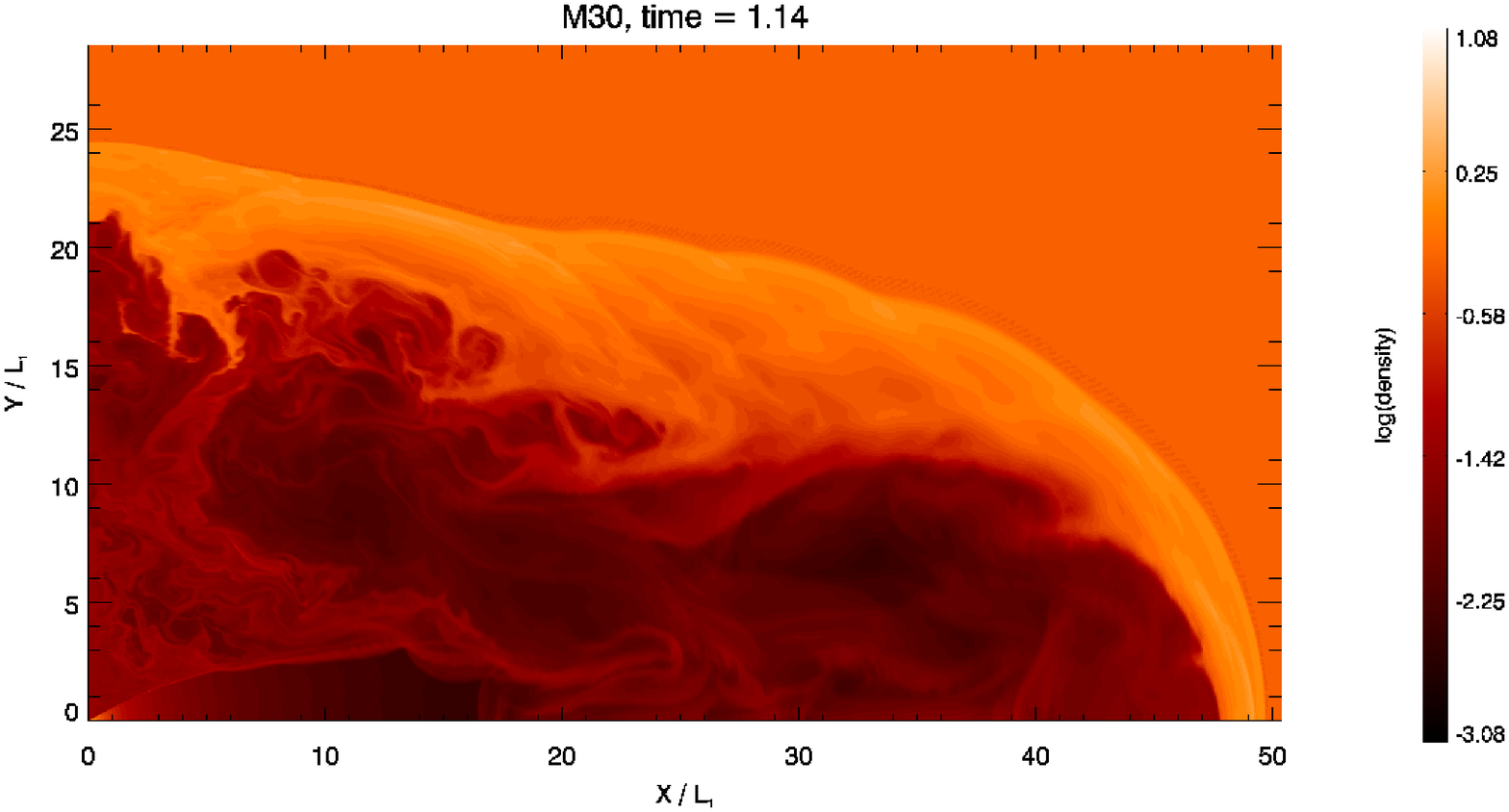} \\
  \vspace*{0pt}
 \caption{Logarithmic density distribution for run M30 for six snapshot times
   indicated in the title of the individual images. }
\label{fig:M30evol}
\end{figure*}
\begin{figure*}
  \centering
  \includegraphics[width=\textwidth]{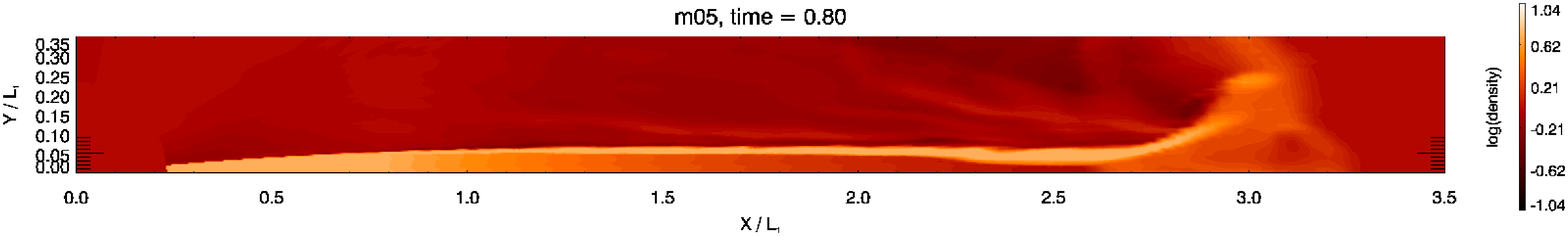} \\
  \includegraphics[width=\textwidth]{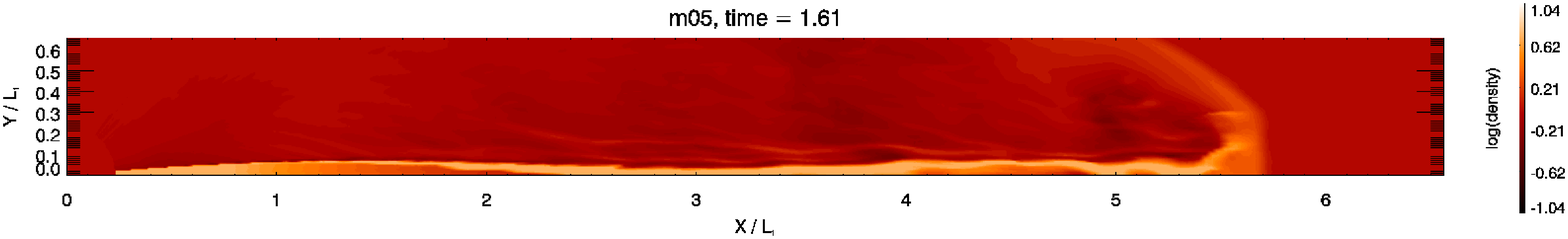} \\
  \includegraphics[width=\textwidth]{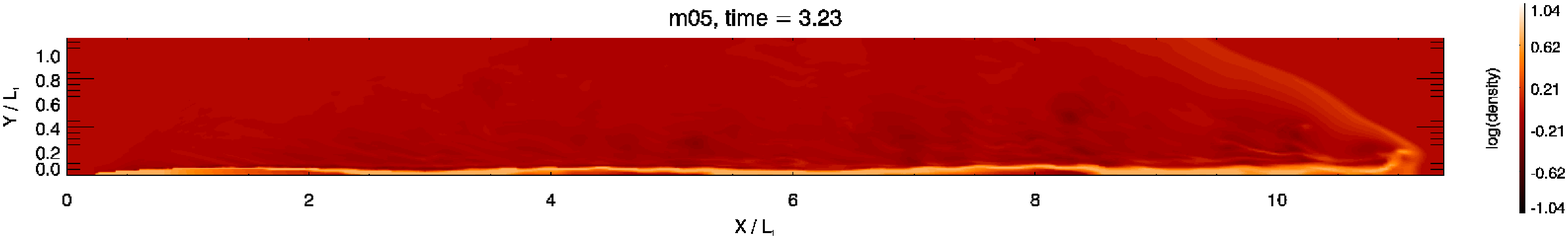} \\
 \vspace*{0pt}
 \caption{Logarithmic density distribution for run m05 for three snapshot times
   indicated in the title of the individual images. }
\label{fig:m05evol}
\end{figure*}
\begin{figure*}
  \centering
  \includegraphics[width=.48\textwidth]{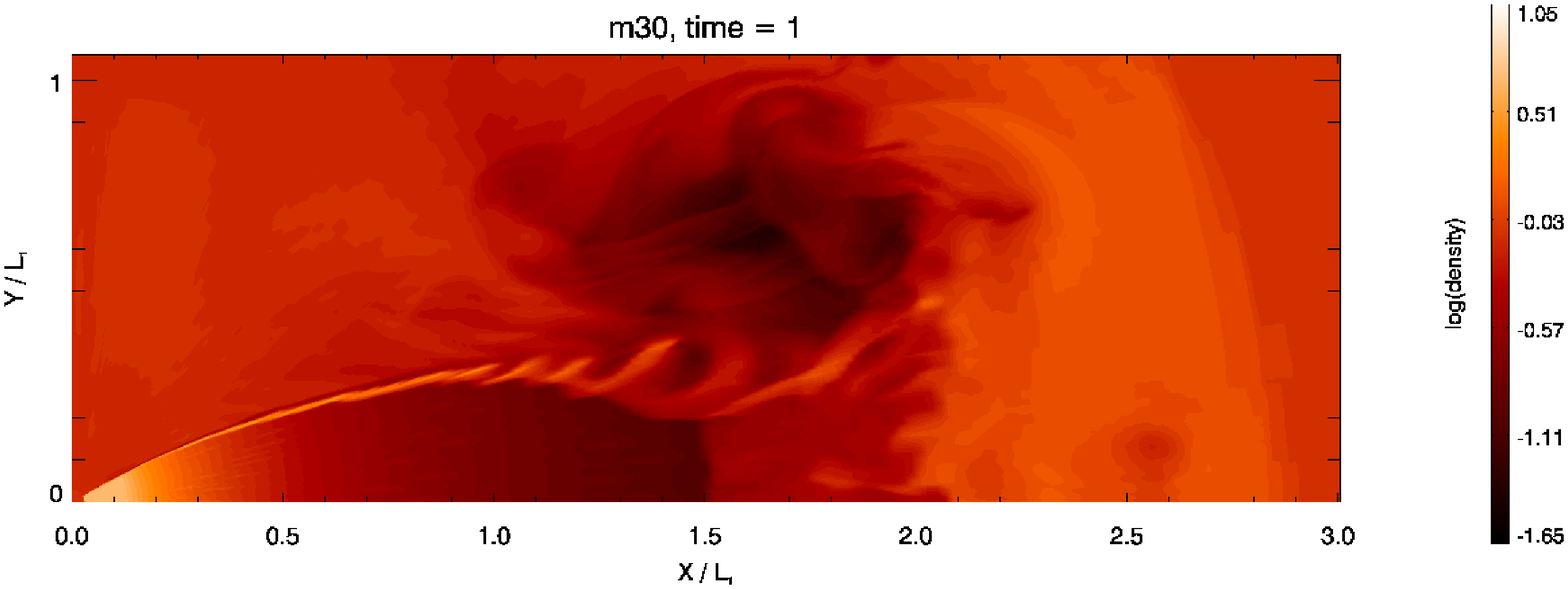} 
  \includegraphics[width=.48\textwidth]{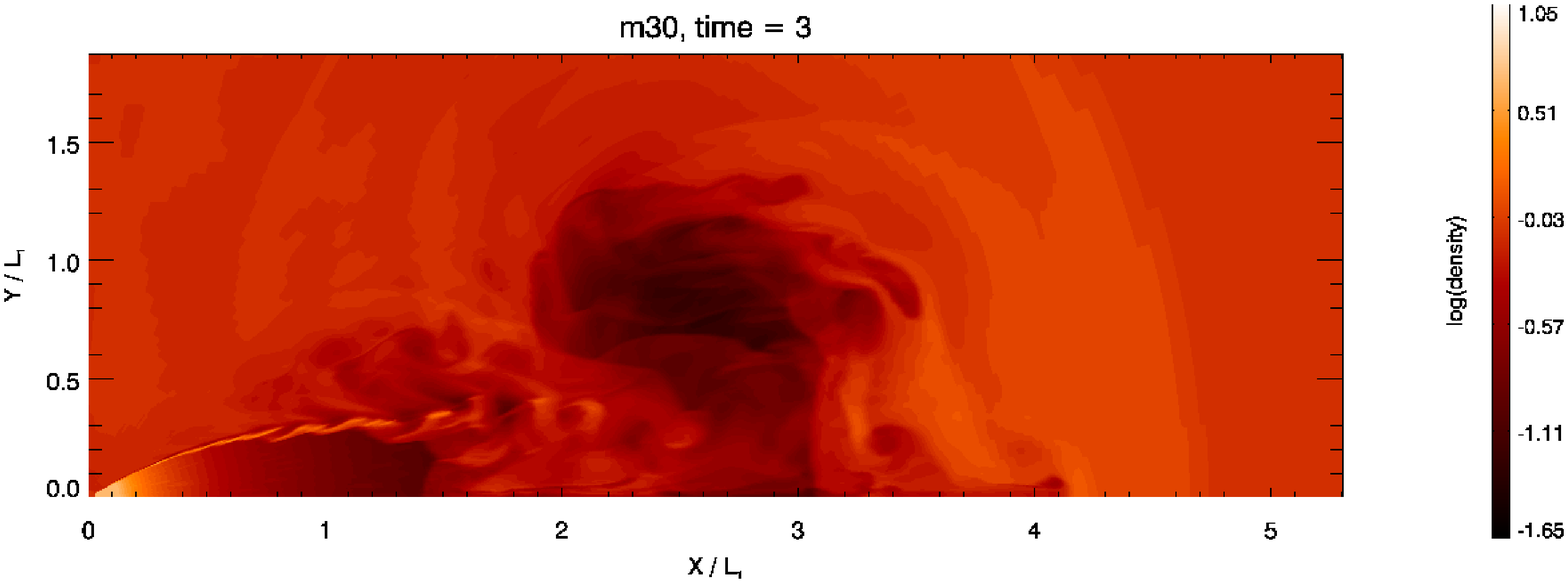} \\
  \includegraphics[width=.48\textwidth]{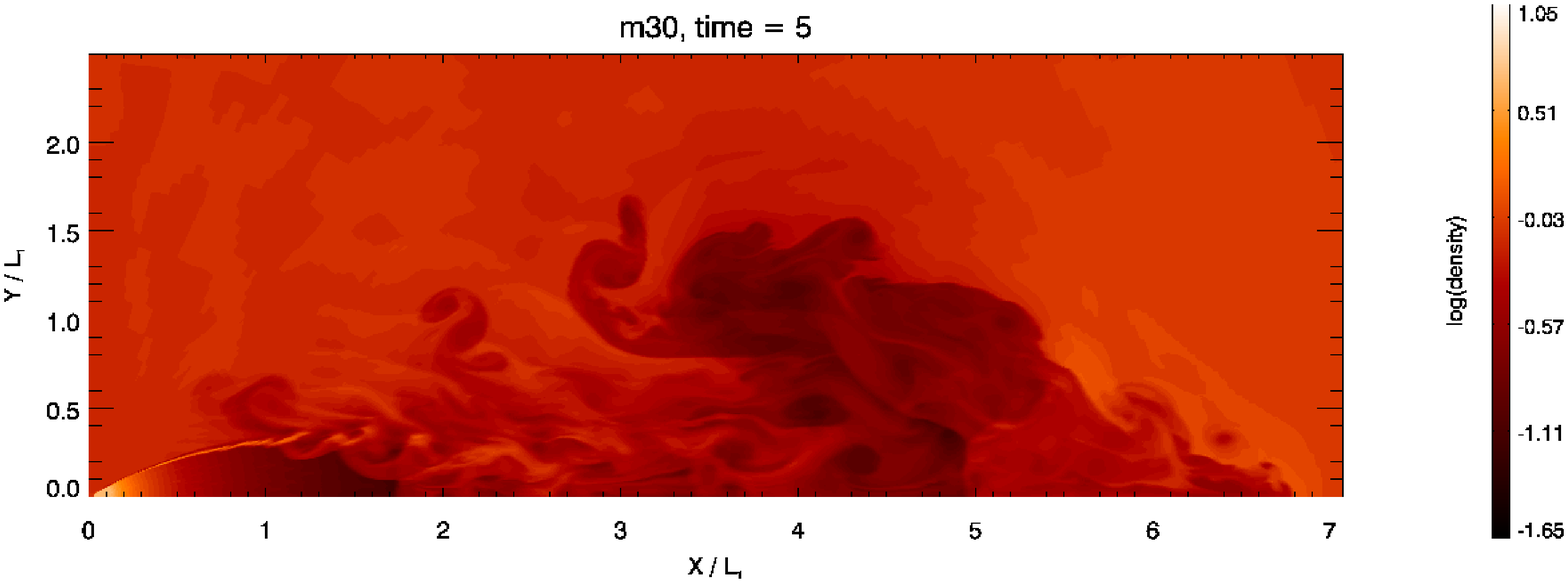} 
  \includegraphics[width=.48\textwidth]{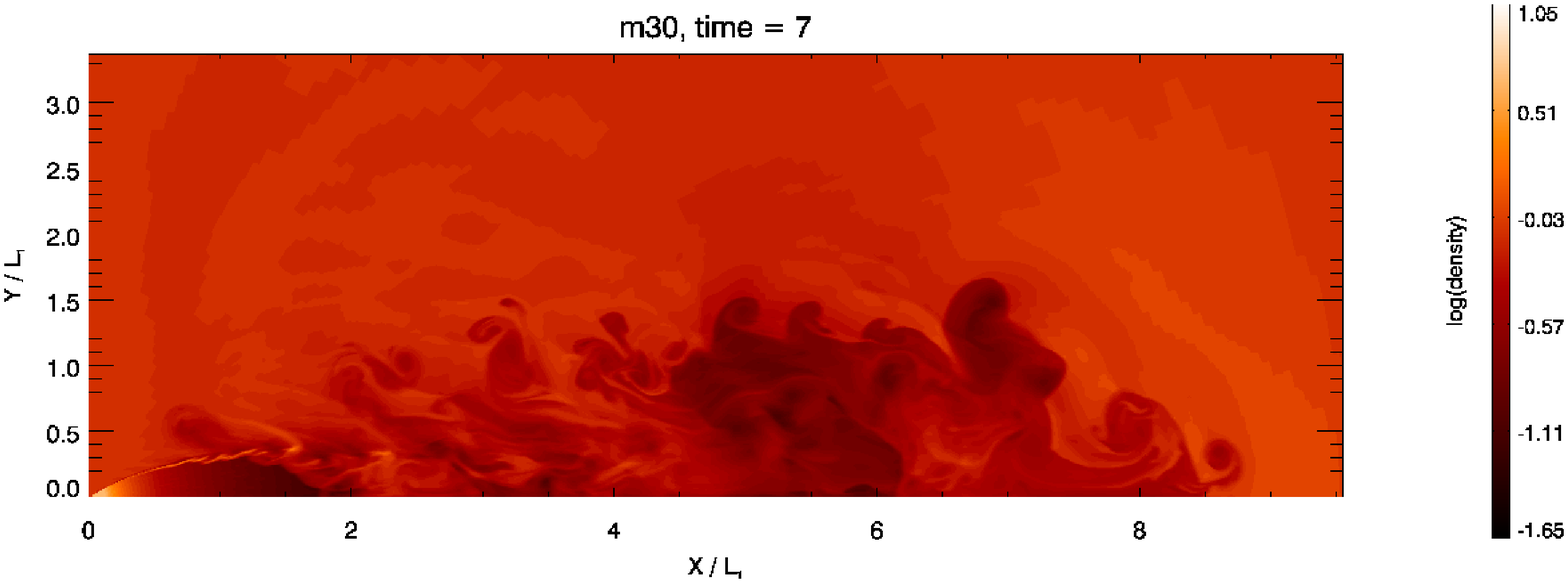} \\
  \includegraphics[width=.48\textwidth]{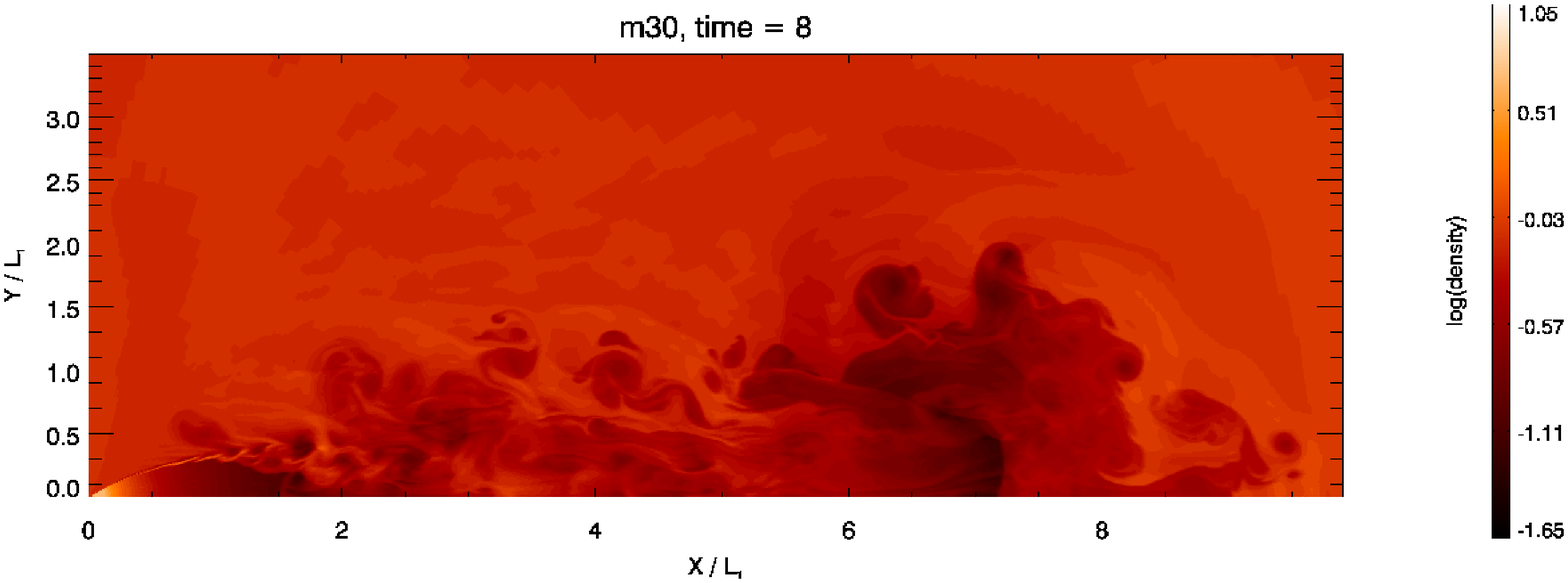} 
  \includegraphics[width=.48\textwidth]{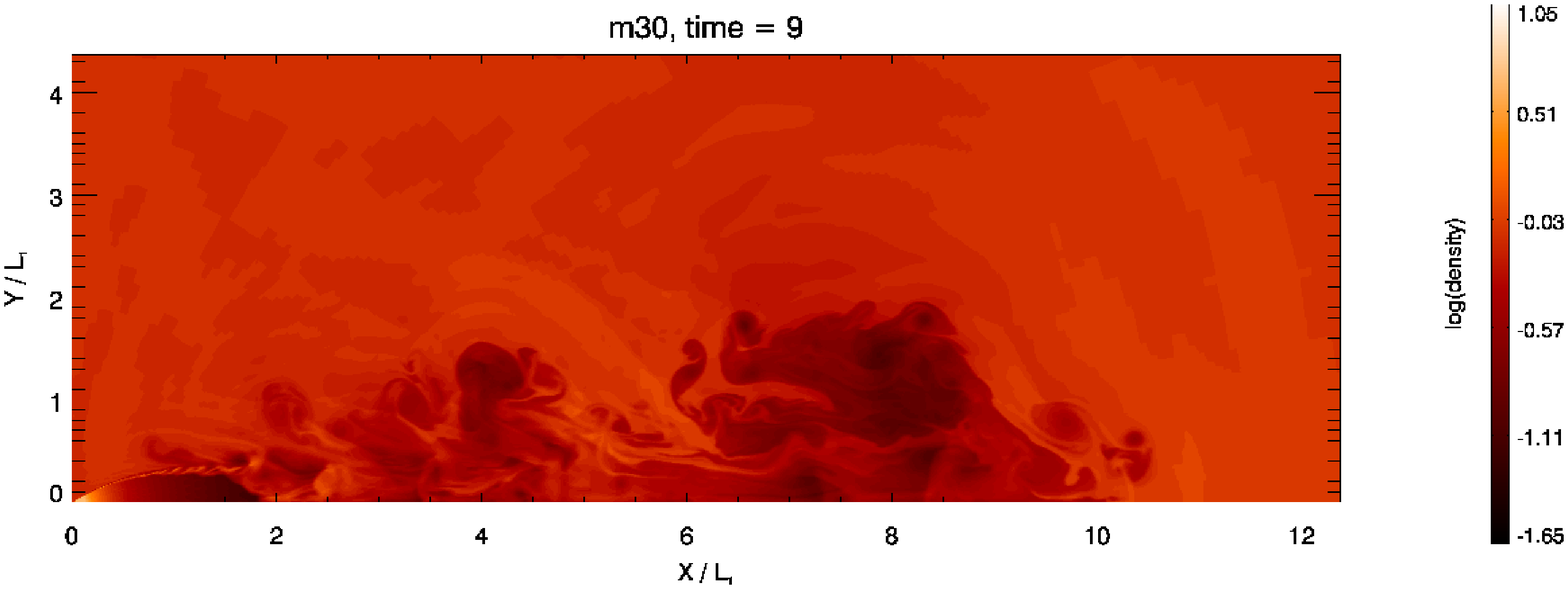} \\
  \vspace*{0pt}
 \caption{Logarithmic density distribution for run m30 for six snapshot times
   indicated in the title of the individual images. Our FR-index is
   above~1.5 for the plots at times 4, 14 and 19, and below 1.5 at
   times 7, 22 and 26.}
\label{fig:m30evol}
\end{figure*}
\begin{figure*}
  \centering
       \includegraphics[width=.48\textwidth]{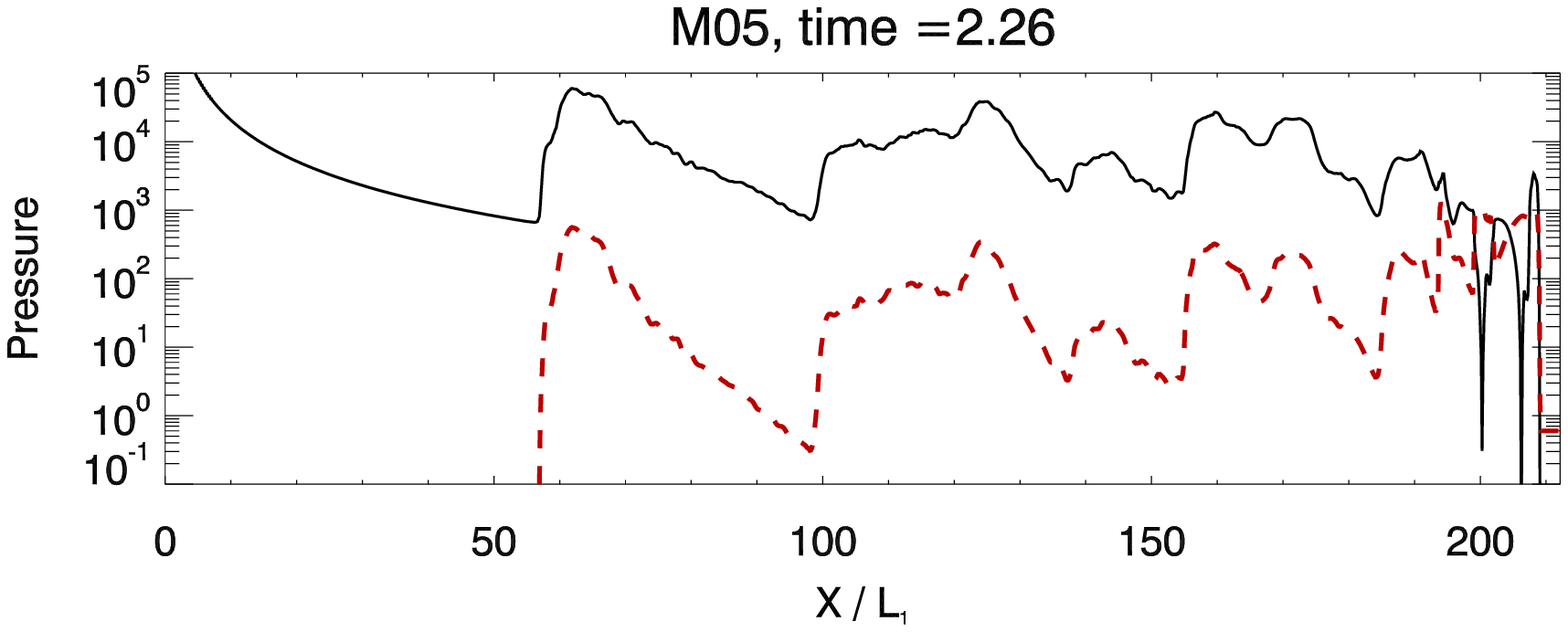} 
       \includegraphics[width=.48\textwidth]{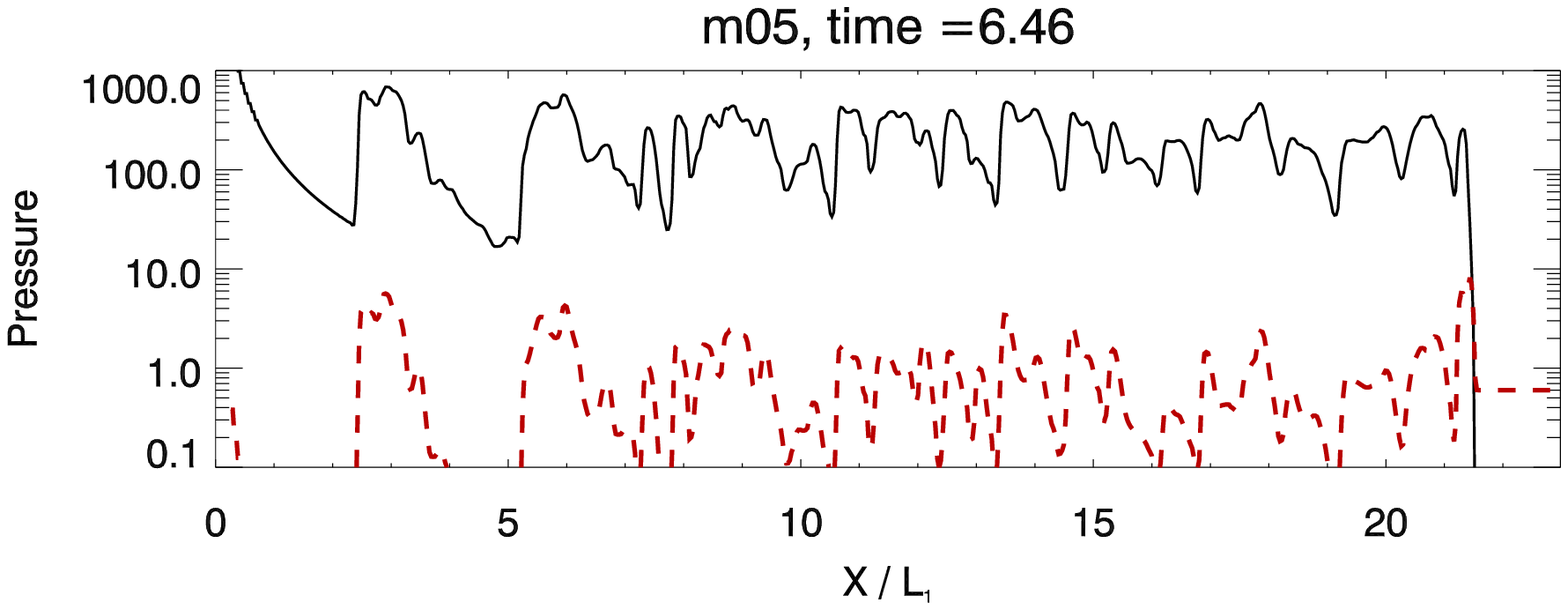} \\
       \includegraphics[width=.48\textwidth]{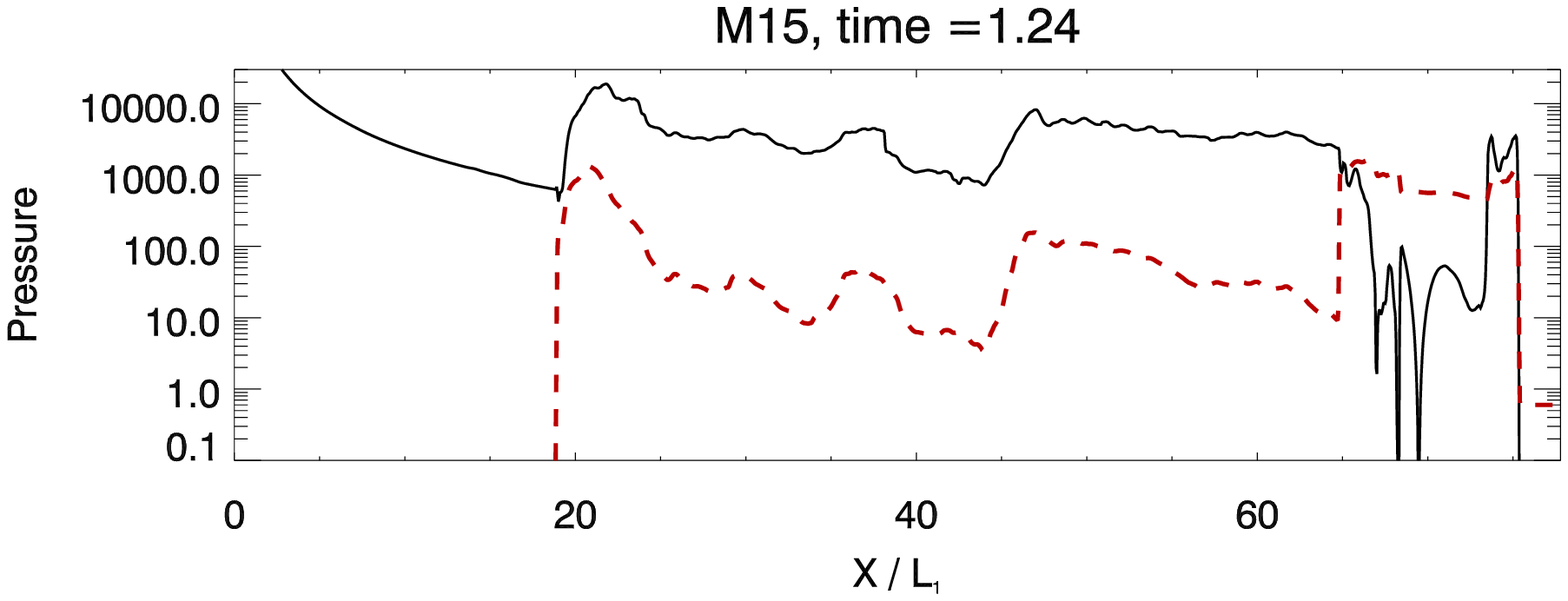} 
       \includegraphics[width=.48\textwidth]{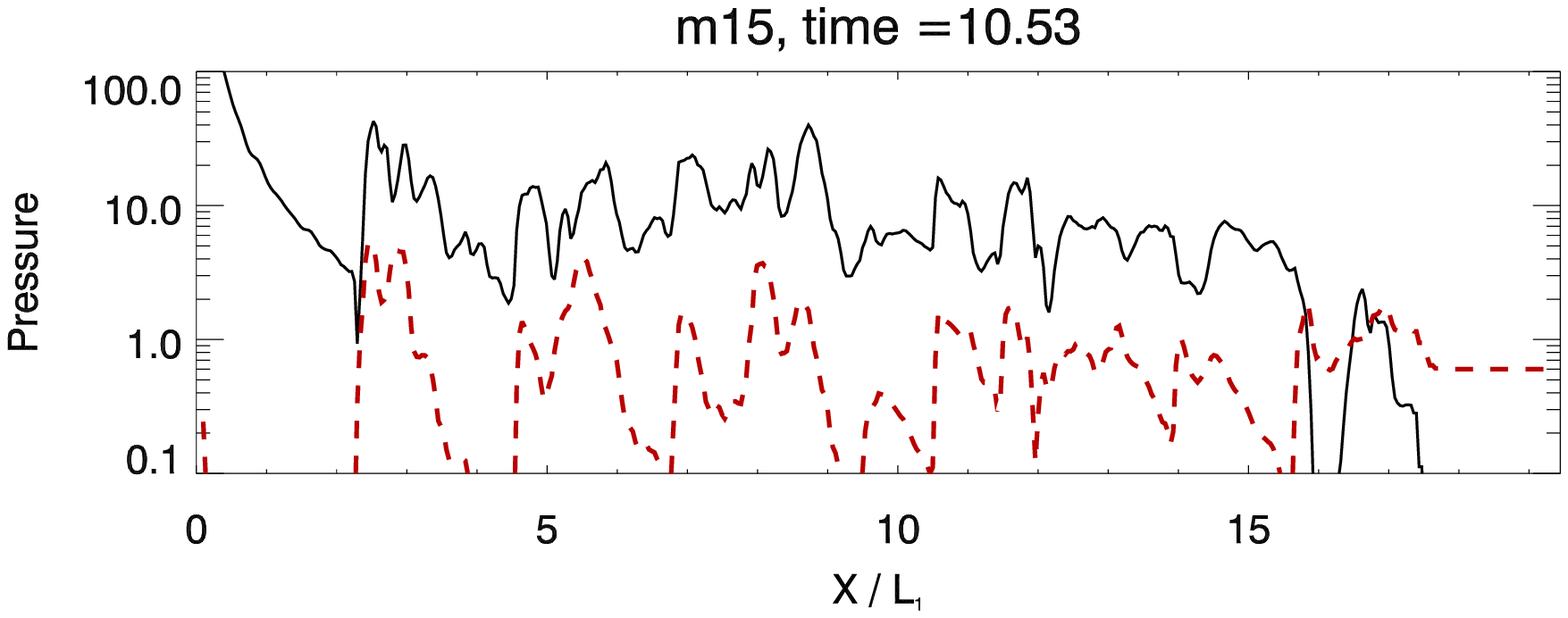} \\ 
       \includegraphics[width=.48\textwidth]{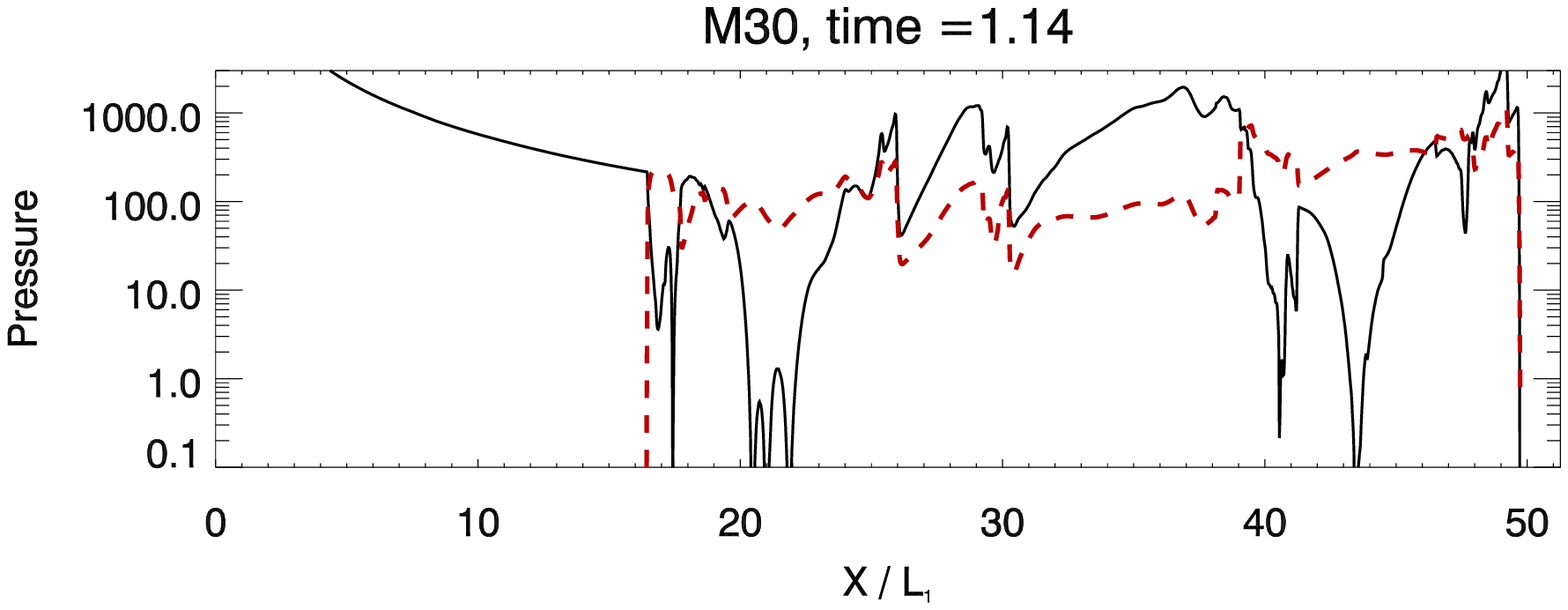} 
       \includegraphics[width=.48\textwidth]{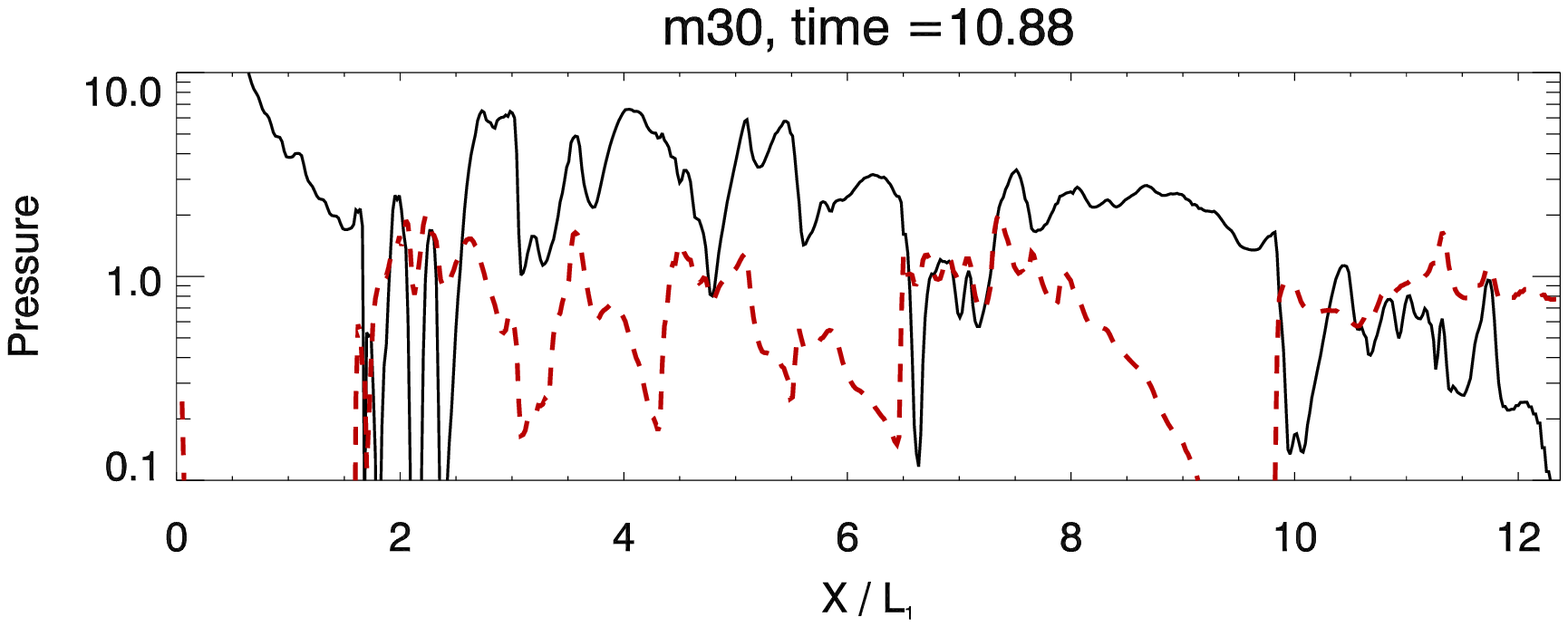} \\
     \vspace*{0pt}
 \caption{Comparison of thermal pressure (red dashed line) against
   forward ram-pressure (black solid line) on
   the axis \change{of symmetry} for \change{the final snapshot of} all simulations.
\cmt{Julia, ``thermal pressure'' is the just the normal pressure. I use the
  adjective ``thermal'' here to discern it from the ram-pressure. The
  simulations are axisymmetric, and therefore have one axis of
  symmetry. The term ``on the axis'' refers to this axis. Does this
  make it clearer?}}
\label{fig:prp}
\end{figure*}
\begin{figure*}
  \centering
  \includegraphics[width=1.10\textwidth]{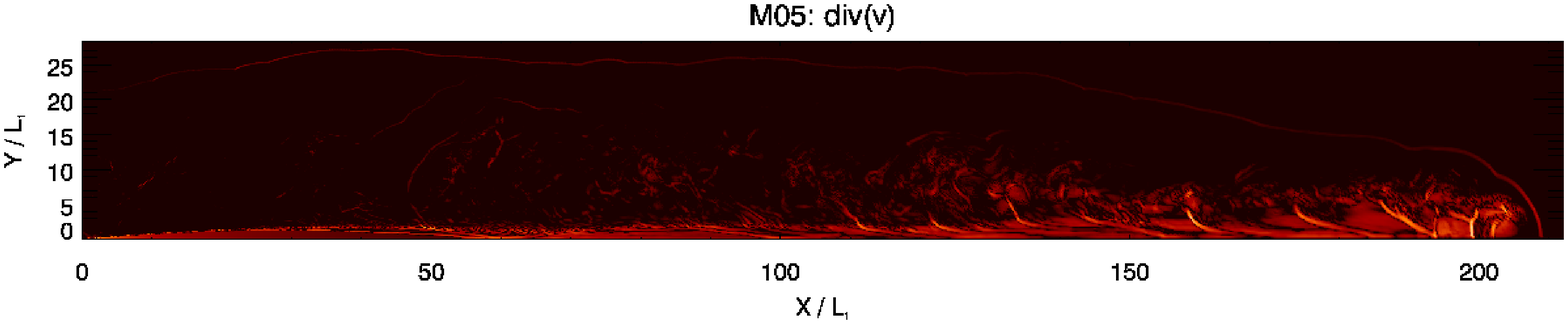} \\
  \includegraphics[width=\textwidth]{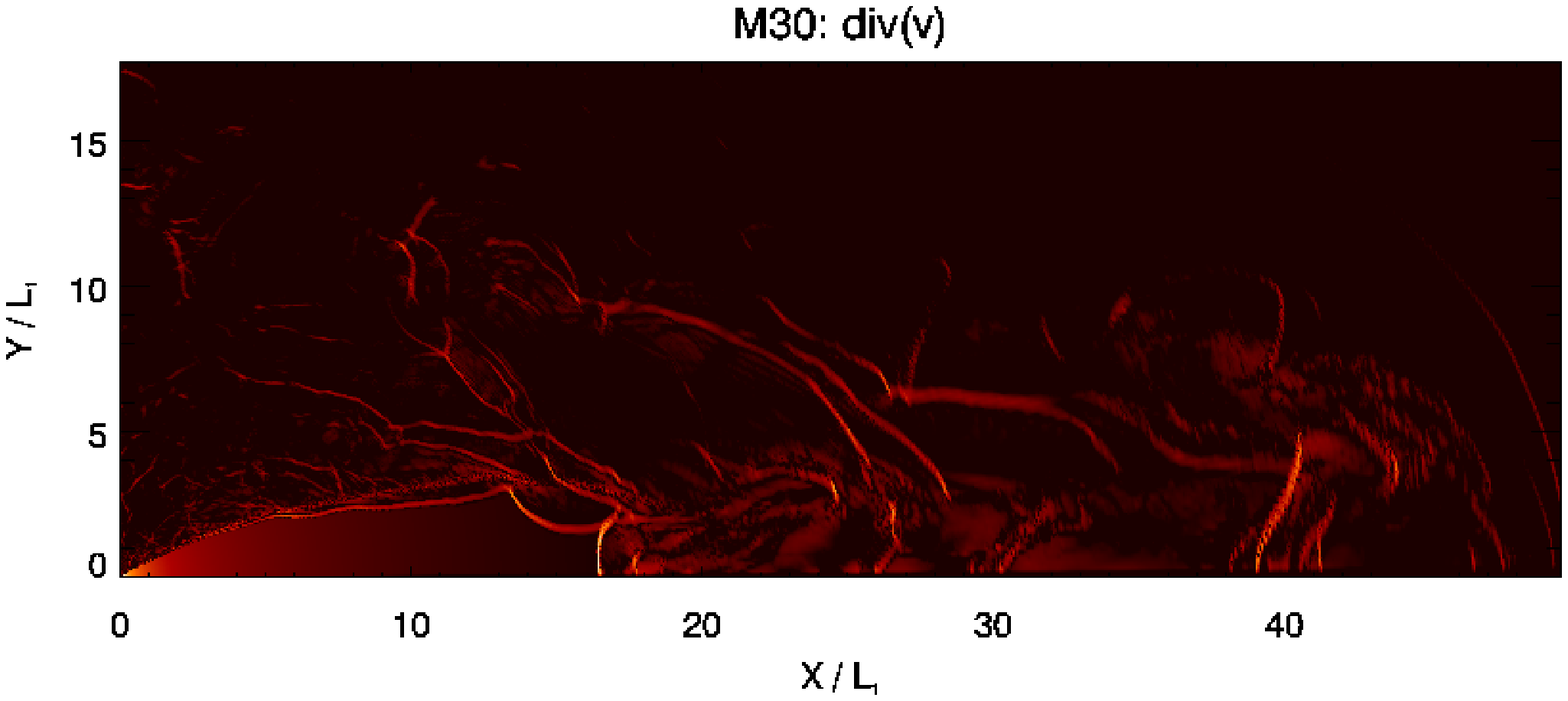} \\
  \includegraphics[width=\textwidth]{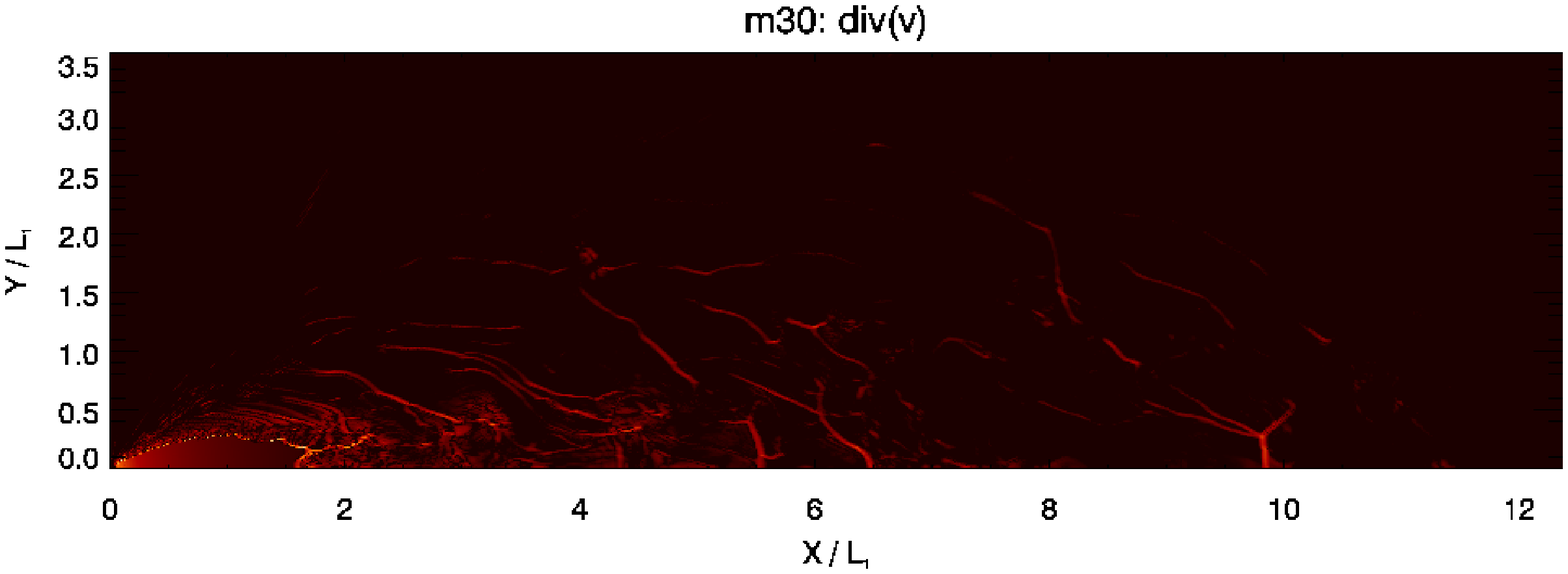} \\
  \vspace*{0pt}
 \caption{Examples of the divergence of the velocity field, shown is
   log($|\nabla \cdot v|$) for runs M05 (top), M30 (middle) and m30
   (bottom). The final snapshot (same as in Figure~\ref{fig:lgd}) 
   is chosen. Shocks are visible as sharp linear features. Adiabatic
   expansion (and compression) appears as diffuse structures. The
   expansion region in the initial conical part shows up clearly.
 \cmt{Figure made bigger.}}
\label{fig:divv2d}
\end{figure*}
\begin{figure*}
  \centering
  \includegraphics[width=.48\textwidth]{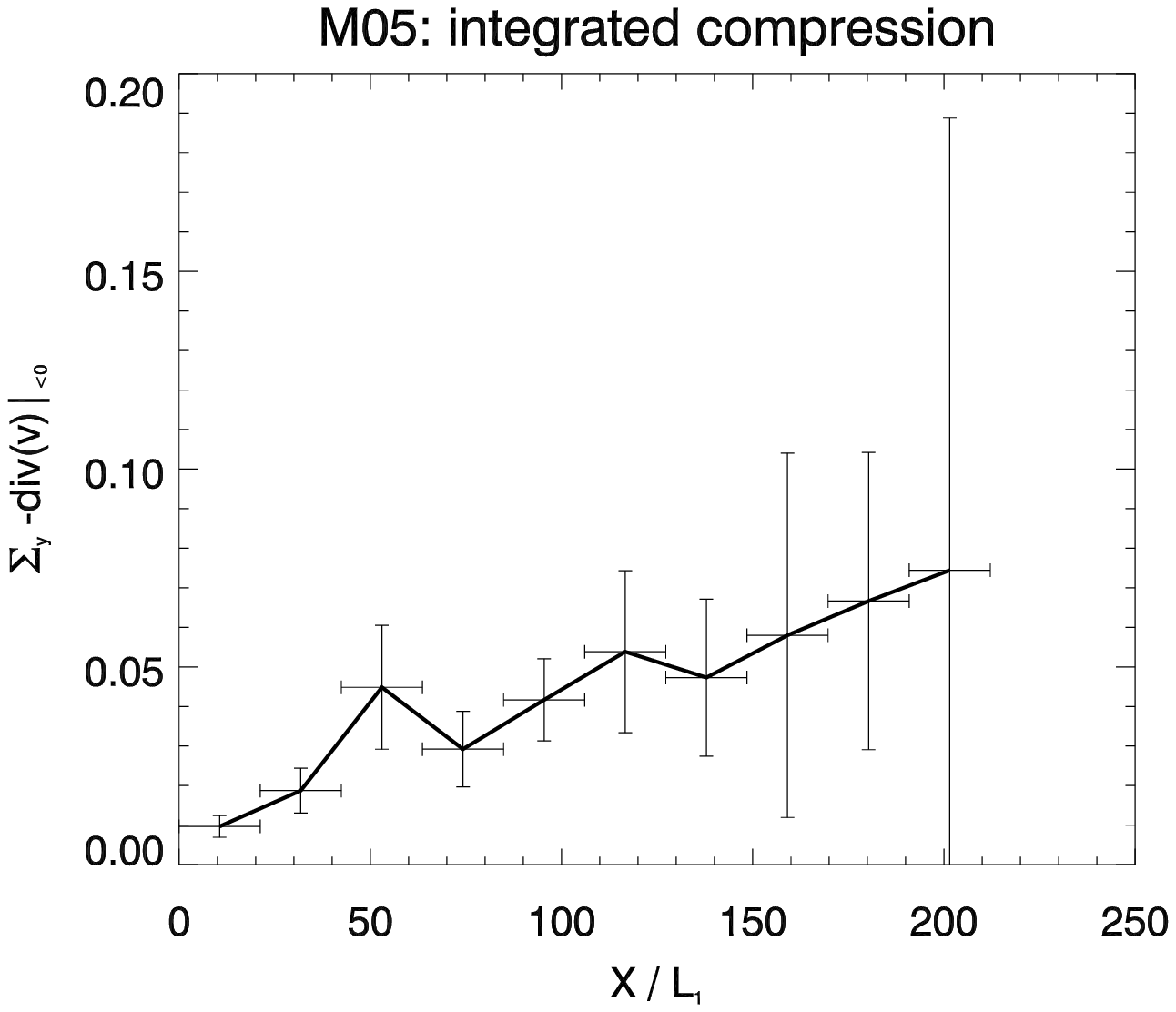} 
  \includegraphics[width=.48\textwidth]{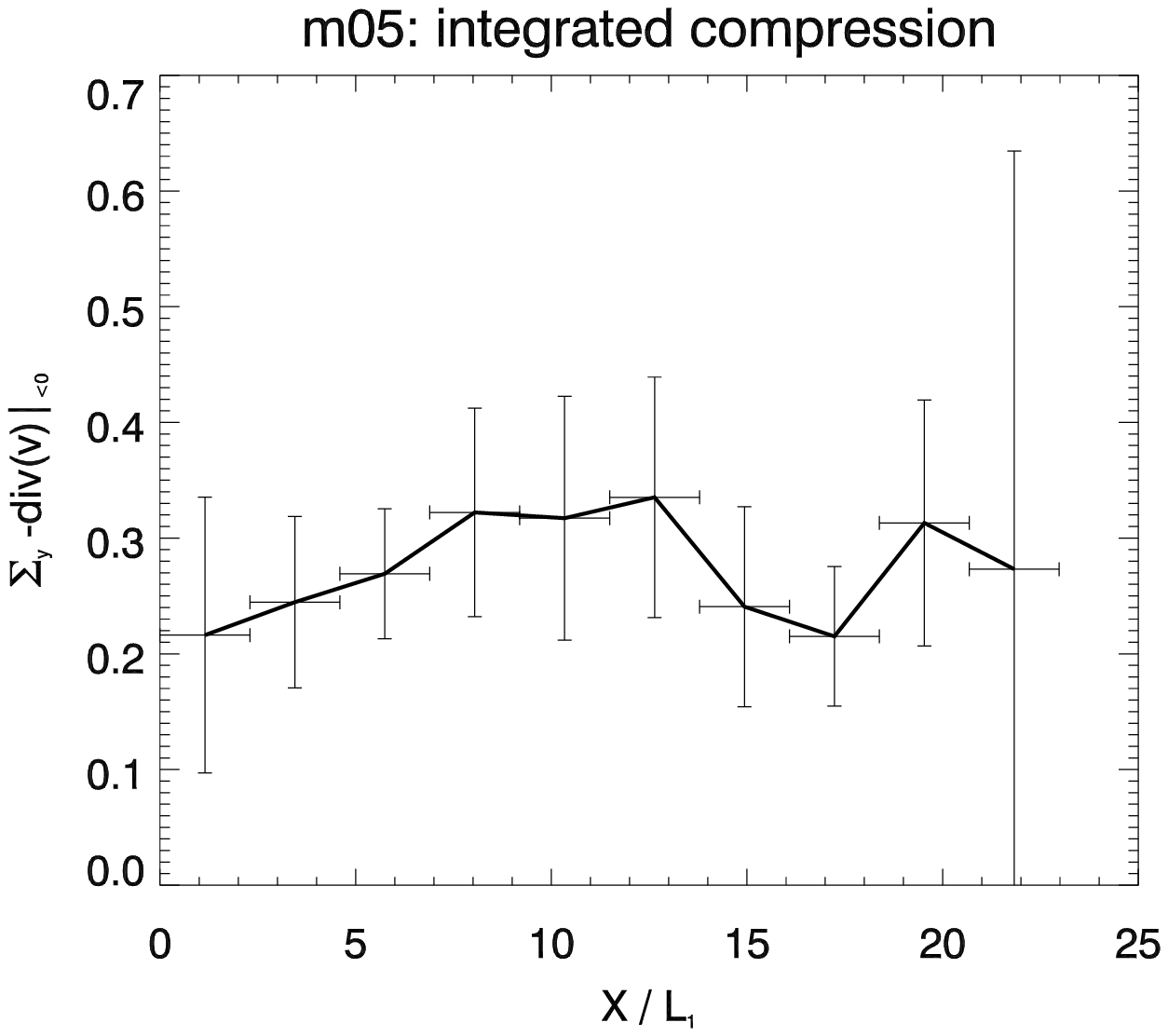} \\
  \includegraphics[width=.48\textwidth]{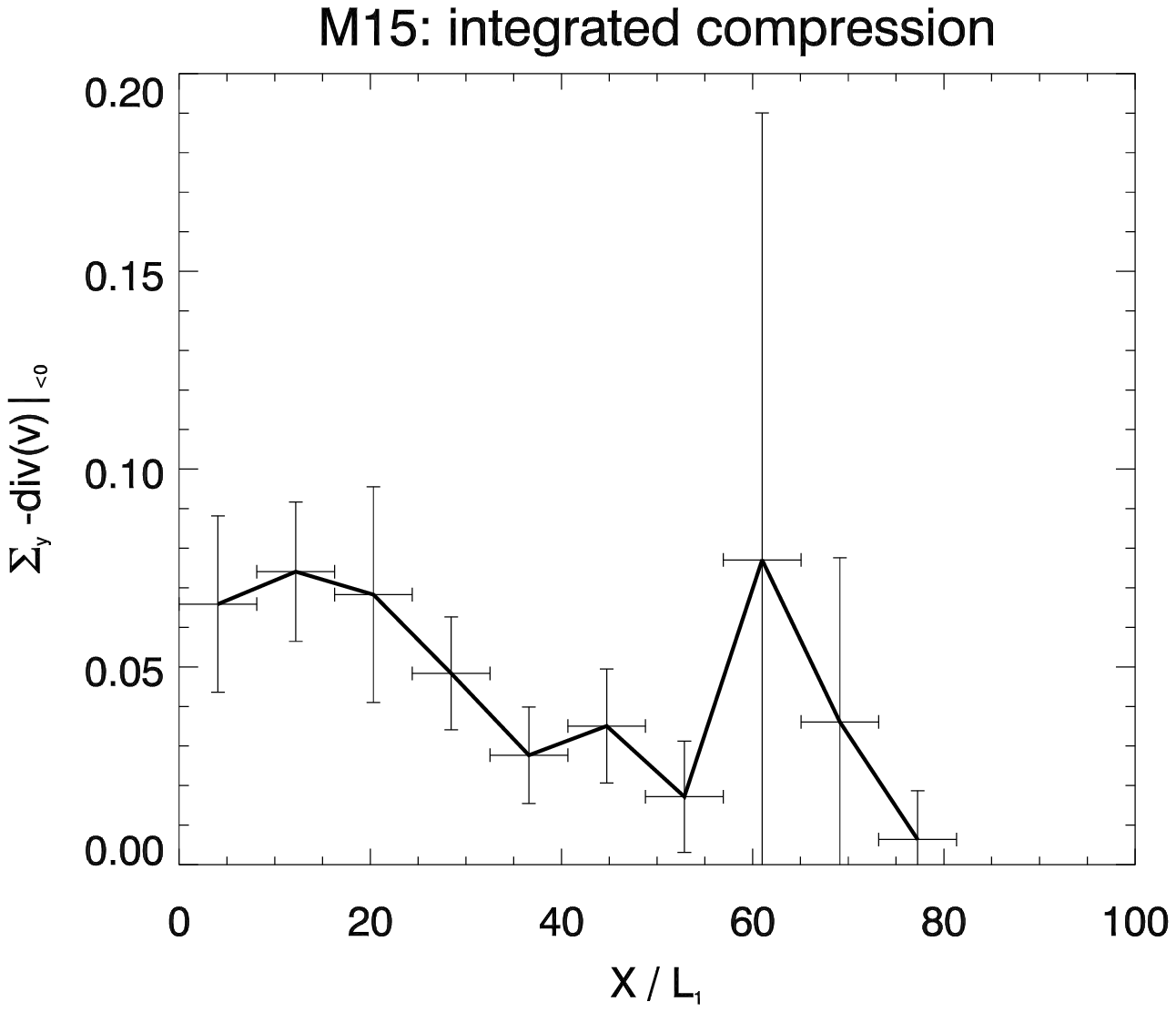} 
  \includegraphics[width=.48\textwidth]{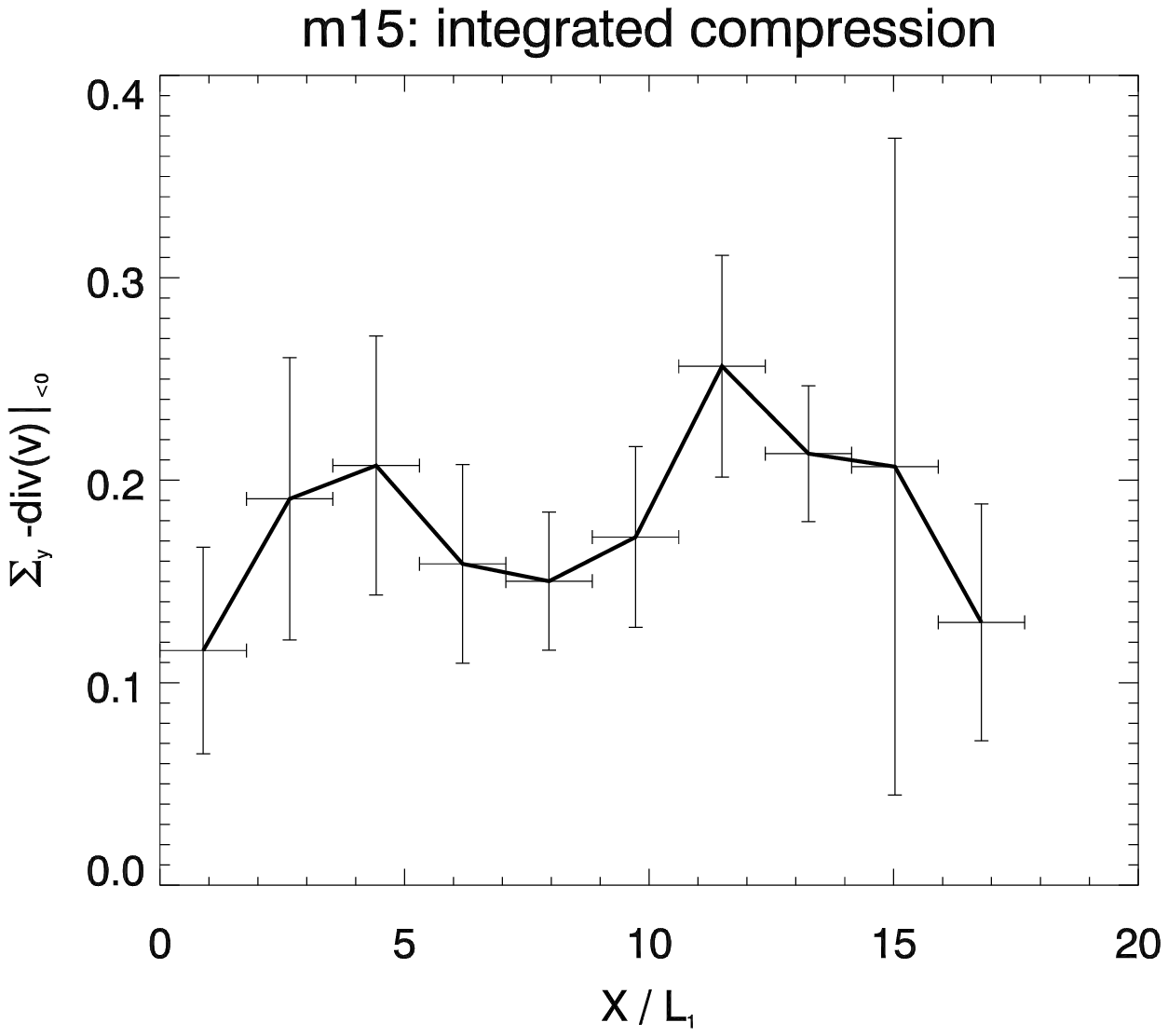} \\
  \includegraphics[width=.48\textwidth]{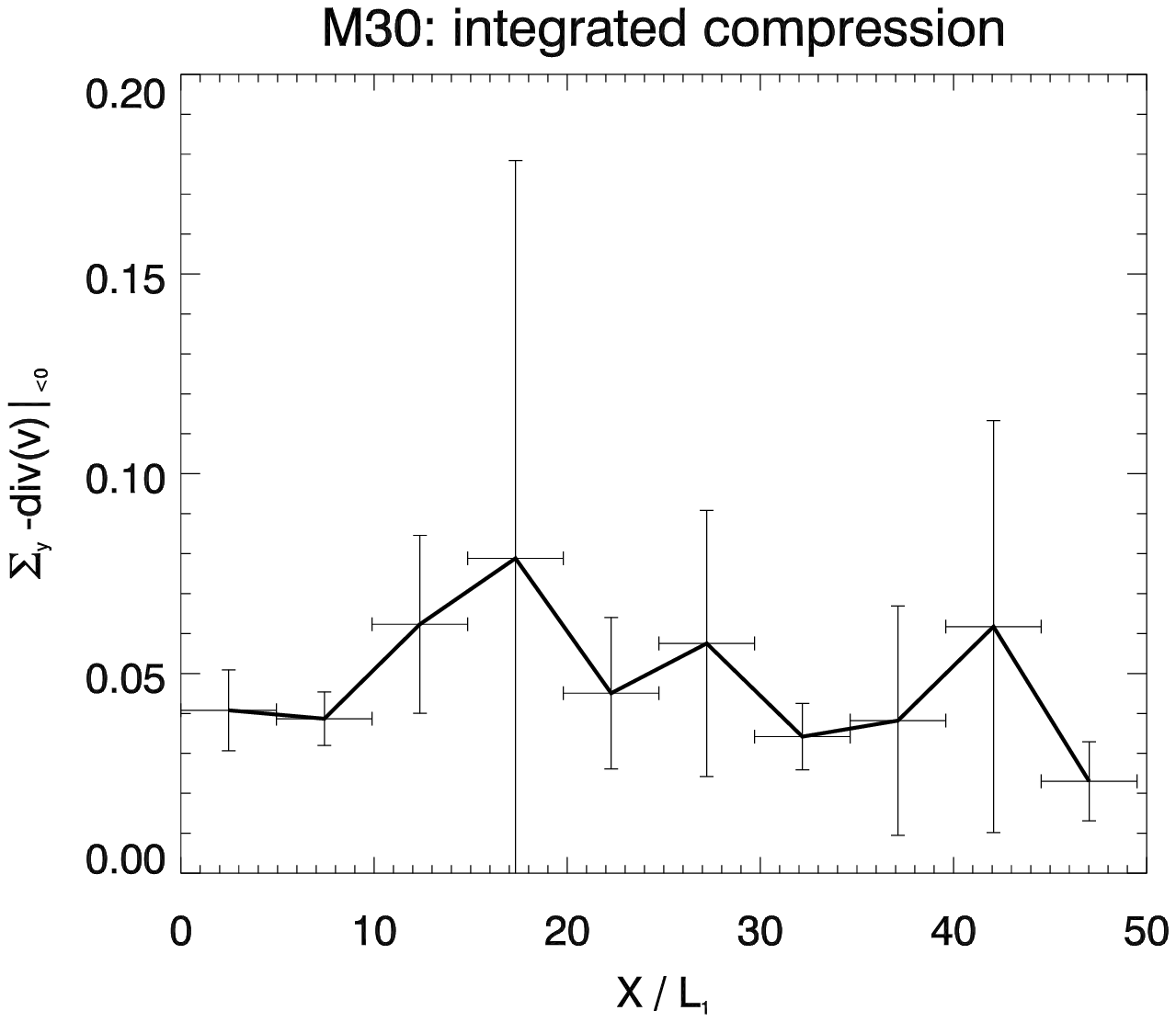} 
  \includegraphics[width=.48\textwidth]{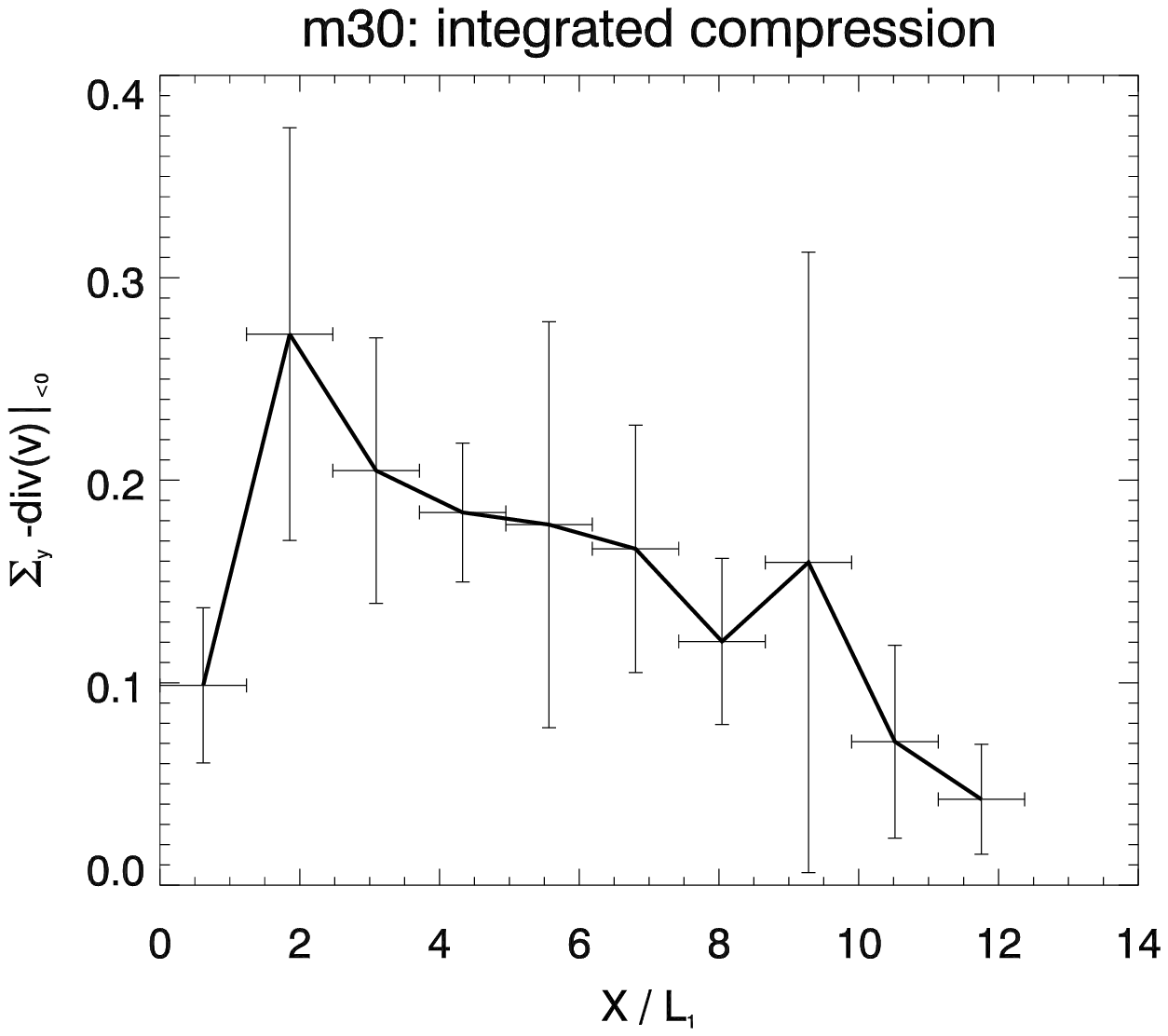} \\
  \vspace*{0pt}
 \caption{Negative part of the divergence of the velocity field
   integrated along a path perpendicular to the symmetry axis,
   i.e. $-\int_y \mathrm{d}y \, \nabla \cdot v|_{<0}$, normalised to
   the respective maximum value. This is a
   measure of the shock strength. The final snapshot is shown for each
   simulation. We use ten bins along the
   $X$-axis. Horizontal error bars are the bin width. Vertical error
   bars are one standard deviation. The regions of the strongest shocks are
   where the error bar of the standard deviation \change{is largest}.}
\label{fig:divv1d}
\end{figure*}

\section{Results}\label{sec:res}
\subsection{Structure}
The spatial structure of the simulated radio sources is shown in
Figure~\ref{fig:lgd}. 
From the discussion above, and
Table~\ref{tab:simpars}, we expect \change{-- and find --} all the
high Mach-number simulations to be
cocoon dominated, since $L_\mathrm{1b}$ is only a few $L_1$, and
$L_\mathrm{1a}$, the length-scale which determines collimation, is
much greater. $L_2$
is 20 to 80 times the maximum jet size, and therefore all the three jets are
overpressured with respect to the external medium throughout the
simulations. The strong bow
shock seen in all three cases is evidence for this
(Figures~\ref{fig:M05evol} and~\ref{fig:M30evol}). Because of the
high cocoon pressure, collimation occurs before
$L_\mathrm{1a}$; there is however reasonable agreement with the length
scales calculated for the sideways ram-pressure equalling the cocoon
pressure. We denote the scales which refer to the cocoon pressure at
the end of the simulations by $x_{1a}$ and $x_{1c}$. They are also
given in Table~\ref{tab:simpars}. \change{Since} the cocoon pressure
  declin\change{es} with time, the re-collimation shock reaches the axis later
  for later simulation times (Figure~\ref{fig:M05evol}\change{,
    $X\approx 100 L_1$ at $t=1.35$, $X\approx 140 L_1$ at $t=2.71$ and
    $X\approx 180 L_1$ at $t=5.42$, note the varying scale on the
    X-axis} \cmt{I added the preceding explanation because Julia noted
  that the effect was hard to see in Figure~\ref{fig:M05evol}}).
The re-collimation shock still reaches the
axis comparatively late in run~M05 -- this is due to a pressure gradient in the
cocoon towards the left (Figure~\ref{fig:lgd}).
These features have already been noted by \citet{KF98}, and we
find good agreement with their simulations.
As predicted by Equation~(\ref{L1ca}), 
the re-collimation shock \change{always} reaches the axis for the
  simulated jets with small opening angles and never \change{reaches
    it} for the ones with
  the large opening angle\change{s}.

All the high Mach-number runs form very light jets and
cocoons with $\eta \approx 10^{-2}$. This is consistent with their 
$L_\mathrm{1b}/x_\mathrm{1a}$ ratios.
At the times shown in Figure~\ref{fig:lgd}, the cocoons are
overpressured by a factor of several tens to hundreds. As the jets
evolve towards $L_2$, the \mghk{cocoon pressure} drops and the \mghk{collimated} jets
become even more underdense.

As predicted in Section~\ref{sec:scales}, \change{for run~M30} 
the terminal shock cannot be
propagated through the re-collimation shock. The jet does not \change{therefore}
re-collimate properly, and a stable collimated beam cannot form. The
terminal shock is however not strictly a standing shock. Its
location is where the forward ram pressure equals the cocoon pressure
($x_{1c}$). At the end of the simulation $x_{1c} = 18 L_1$, as
calculated from the current cocoon pressure. This is in good agreement
with the measured position. As can be seen from
Figure~\ref{fig:M30evol}, the position of the terminal shock advances
with time, according to the declining cocoon pressure. \change{I}n our
simulation\change{,} the terminal shock \change{never} advance\change{s} beyond the
re-collimation shock\change{,} as expected from Equation~(\ref{L1ca}) \change{--} the
ratio between the terminal shock limit and the re-collimation scale
only depends on the initial opening angle\change{;  r}un~M30 demonstrates
this result for a wide range of cocoon pressures.

The low Mach-number runs may be found in the right column of
Figure~\ref{fig:lgd}. In run~m05, the re-collimation shock reaches the
axis at about~3~$L_1$, consistent with its $L_\mathrm{1a}$ value. This is
only slightly larger than $L_\mathrm{1b}$, and hence essentially no
cocoon is formed. \change{W}e expect \change{-- and find --} 
the first reflection point of
the re-collimation shock, where it meets the axis, to have a constant
$x$-value throughout the simulation
(Figure~\ref{fig:m05evol}). The density structure of the beam is a
hollow spine with a dense sheath. On average, the beam stays overdense
with respect to its environment, as expected from Equation~(\ref{eta}). A
minor filamentary cocoon forms for run~m15
\change{(Figure~\ref{fig:lgd})}. The beam density in
this case is almost matched to the environment. This is also as
expected from the values of $L_\mathrm{1a}$ and $L_\mathrm{1b}$ 
(Table~\ref{tab:simpars}). For run~m30,
$L_\mathrm{1b}/L_\mathrm{1a}=0.3$, and we expect an underdense jet
with a significant cocoon, which is confirmed by the simulation.
All the low Mach-number jets have expanded considerably \pa{beyond
$L_2$ by the end of the simulation.} Consequently, very weak bow shocks, if any at all, are found,
except near the tip of the supersonically propagating runs~m05 and m15.
In a similar fashion to the high Mach-number case, only the simulation with a
half opening angle of $30^\circ$ develops a stationary terminal
shock (Figure~\ref{fig:m30evol}). The predicted position ($L_\mathrm{1c}$ in
Table~\ref{tab:simpars}) is $3.5L_1$. From the simulation, we measure a
value of $1.7L_1$, and it oscillates between $1.4L_1$
and $1.8L_1$. 
The reason
for the slight disagreement is the pressure enhancement 
downstream of the shock (Figure~\ref{fig:prp}).

The cocoon in run~m30 has transformed in an interesting way
(Figure~\ref{fig:m30evol}).  While for the other simulations, the
cocoon is oriented backwards, surrounding all of the jet, here it
transforms into a roughly conical forward flowing structure. 
\change{Compared with}~M30, \change{in which} the cocoon
\change{is overpressured by} a factor of $\approx 100$, 
m30 has expanded to more than three
times $L_2$ and is in pressure equilibrium with the
environment. Hence, if a cocoon \change{had} formed around the inner
jet, it would have become Rayleigh-Taylor unstable.
We have argued in Section~\ref{sec:scales} that such a morphological
transformation should be related
to strong entrainment.
The re-collimation shock is unstable in the high opening angle runs;
\change{additionally in run~m30} 
the ambient gas is in direct contact with the jet \change{and} efficient
entrainment \change{indeed}  happens at the re-collimation shock.

The terminal shock
limit should be determined by the forward ram pressure becoming equal to the
local thermal pressure. We confirm this from our simulations in
Figure~\ref{fig:prp}. For the runs with half opening angle $30^\circ$,
where the terminal shocks occur before re-collimation
there is indeed a close match between the two near their terminal shock
regions ($X=1.7 L_1$ and $X=17 L_1$ for runs m30 and M30, respectively). In run~M30,
the terminal shock is clearly kept at its position by the cocoon
pressure. 
For the other
simulations, the ram-pressure gets \change{sometimes} 
remarkably close 
\cmt{Julia, you noted ``?factor of 10?'' here. In Figure 6 you have to
compare red and black line at the first maximum of the red
line. Except m05, the lines are almost identical there (M05,M15,m15).}
to the thermal
pressure near the position where the re-collimation shock hits the
axis. It is, however an oblique shock, and no exact match is
required. The internal jet Mach-number is related to the ratio of the
ram pressure to the thermal pressure. From  Figure~\ref{fig:prp}, the internal
Mach-numbers are therefore of order ten for runs~m05, M05~and~M15. For
run~m15, it varies between two and six. Even for the high opening
angle runs, the internal Mach-number downstream of the terminal shock
quite often exceeds two.

\subsection{Fanaroff-Riley classification}
The simulations we present above show a clearly distinct
morphology. In order to link them to \change{the} observed 
\change{structures of} radio sources, we have
to derive properties of the associated synchrotron emission, which is
due to non-thermal\change{, relativistic} particles and magnetic
fields. 
\change{Neither} is included
in our simulations. For the FR classification, we concentrate on low
frequency radio emission. The latter is however commonly
approximated by $j \propto p^{1.8}$, where $p$ is the thermal pressure
\citep[e.g.][]{Saxea02}. 
\change{The idea behind this is the following:
The emitting particles are accelerated in shocks, which also raise
the pressure. After acceleration, the particles suffer adiabatic
losses, while being advected away from the shock. Similarly, the
pressure in the hydrodynamic simulation declines when the plasma
re-expands behind a shock wave. The exponent follows from the equipartition
argument.}

For our low opening
angle runs, strong shocks map on to
pressure enhancements. \change{However, f}or the high 
opening angle runs (M30, m30),
the strongest shock for much of the simulation time
is the \change{standing terminal} shock \change{in the re-collimation region}, but the
pressure is fairly uniform from the post-shock region, downstream into
the cocoon (Figure~\ref{fig:prp}). \change{I}t is \change{therefore} more realistic to
include a measure of the shock strength directly into the emissivity.
The compression rate in the flow is characterised by $\nabla \cdot
{\bf v}$,
where $\bf v$ is the velocity vector. It also traces adiabatic
compression and expansion, but the compression is strongest in shocks.

\begin{figure*}
  \centering
       \includegraphics[width=.999\textwidth]{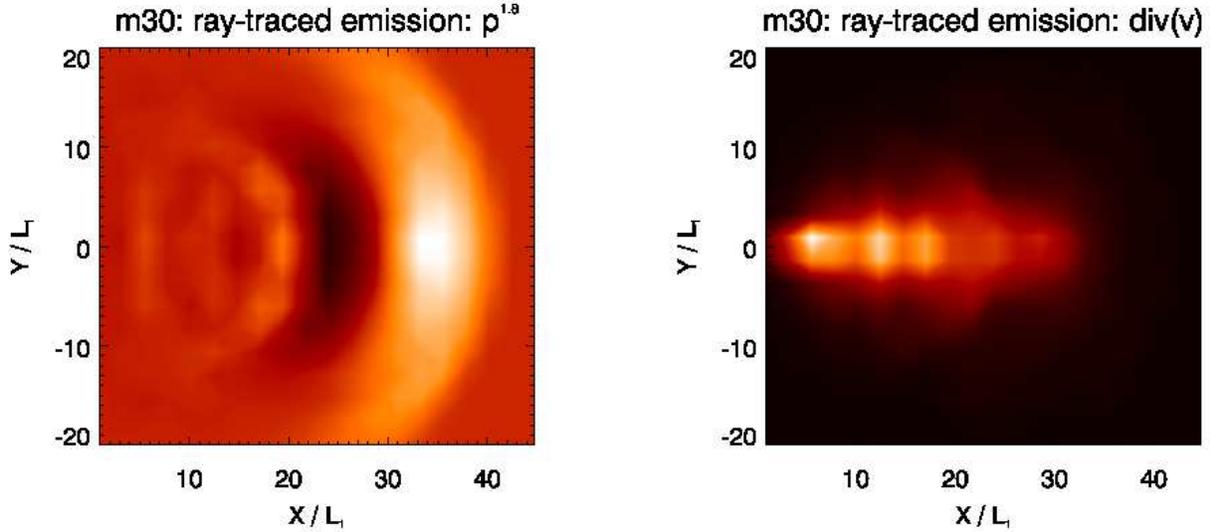} 
    \vspace*{0pt}
 \caption{Line of sight (perpendicular to the jet axis) integrated, 3D
 transformed emissivity for different emission
 tracers. 20 resolution elements are used for the X-axis.
$p^{1.8}$ (left) badly reproduces strong shocks. 
$|\nabla \cdot {\bf v}|$ (right) highlights shocks much better. 
We use the latter for comparison to observations. See text for details.}
\label{fig:emtest}
\end{figure*}

Figure~\ref{fig:divv2d} shows $\nabla \cdot {\bf v}$ for runs M05, M30
and m30 for the final snapshots. Besides the initial adiabatic
expansion in the conical part of the jet, the plots are dominated by linear
shock features. Clearly, for run M05, the strongest shock is the
classical Mach disk near the head of the source. Moderately strong
shocks are found around the Mach disk and somewhat upstream. The
shocks are concentrated in the out\change{ermost} half of the jet.
This is in sharp contrast to the high opening angle runs M30 and m30.
Here, the strongest shock is the \change{standing terminal} shock 
\change{in the re-collimation region}, and many weaker
shocks are located in the vicinity. We quantify this in
Figure~\ref{fig:divv1d}, where we plot the integral of the compression
rate \change{perpendicular to the axis}, 
$-\int_y \mathrm{d}y\,\nabla \cdot {\bf v}|_{<0}$ (i.e. positive
values which are not related to shocks are set to zero). 
We use ten bins along each jet in order to
avoid \change{overemphasising} isolated features, and give the
bin size and the standard deviation of the compression rate in each
bin as horizontal and vertical error bars, respectively. 
All runs with half opening angle of
$15^\circ$ or less display the strongest shocks in the head region,
which is evident from the large error bars there. 
\change{In contrast, f}or the high opening
angle runs the 
\change{standing terminal shock in the re-collimation area} 
is the strongest feature, also, when averages over larger areas 
are considered, as we have done here. 
\change{Correspondingly, the compression rate reaches the highest
  values near that site, in the innermost part of the jet.}
Thus, the shock structure in high opening
angle runs is clearly distinct from the one in low opening angle
simulations. 
\change{Yet, the maximum of the pressure is always close to the
tip of the jet, also for runs~M30 and~m30. 
We exemplify this for the final snapshot of run~m30 in
Figure~\ref{fig:emtest}, where we compare emission maps using
$p^{1.8}$ and $\nabla \cdot {\bf v}|_{<0}$ as respective emission tracers.
Clearly, the assumption
that pressure maxima trace strong shocks is not generally correct for our
high opening angle runs, and we need another proxy for the emissivity.}

\change{Shocks are most directly traced by the compression rate $\nabla \cdot
{\bf v}$. Realistically, the particles would suffer losses while
streaming away from the shocks, which smooths out the emission over a
certain length behind the shock. We cannot hope to model the detailed
appearance of these emission features with our simulations in any
useful way. However, the total emission initiated by any given shock
should be related to the strength of the shock and its area.}

\change{F}or connecting our results to
observations \change{we therefore believe it is appropriate} 
to take the integrated compression rate defined above ($-\int_y
\mathrm{d}y\,\nabla \cdot {\bf v}|_{<0}$) as a measure of the
emissivity\change{. A}t the same time \change{we have}
reduce\change{d} the spatial resolution so that
the emission due to each shock is assigned to a larger volume.

\change{Also in observations the radio emission appears knotty. A given radio
knot is most likely retaled to one particular shock or shock complex.
Additionally, the FR classification has originally been established
using low resolution radio maps \citep{FR74}. It is therefore
appropriate that we also smooth the simulation results to a comparable resolution.}


With this choice for the emissivity, we produced
pseudo-radio maps for all the 2949 snapshots of our six runs, and
smoothed it to 20 resolution elements along the axis. \change{W}e defined
the source size by the extent of the pseudo-radio emission along the
$X$-axis, up to the point where the emission drops below five per cent of
the maximum value ($x_\mathrm{size}$). We also
calculate the position of the brightest feature
$x_\mathrm{bright}$. Finally, we define the FR-index as:
\change{
\begin{equation}
\mathrm{FR} =  2 x_\mathrm{bright} / x_\mathrm{size}\, +1/2.
\end{equation}}
With this definition, \change{$0.5<$FR$<1.5$} for an FR~I
radio source, and \change{$1.5<$FR$<2.5$} for an FR~II radio source.

\begin{figure*}
  \centering
       \includegraphics[width=.48\textwidth]{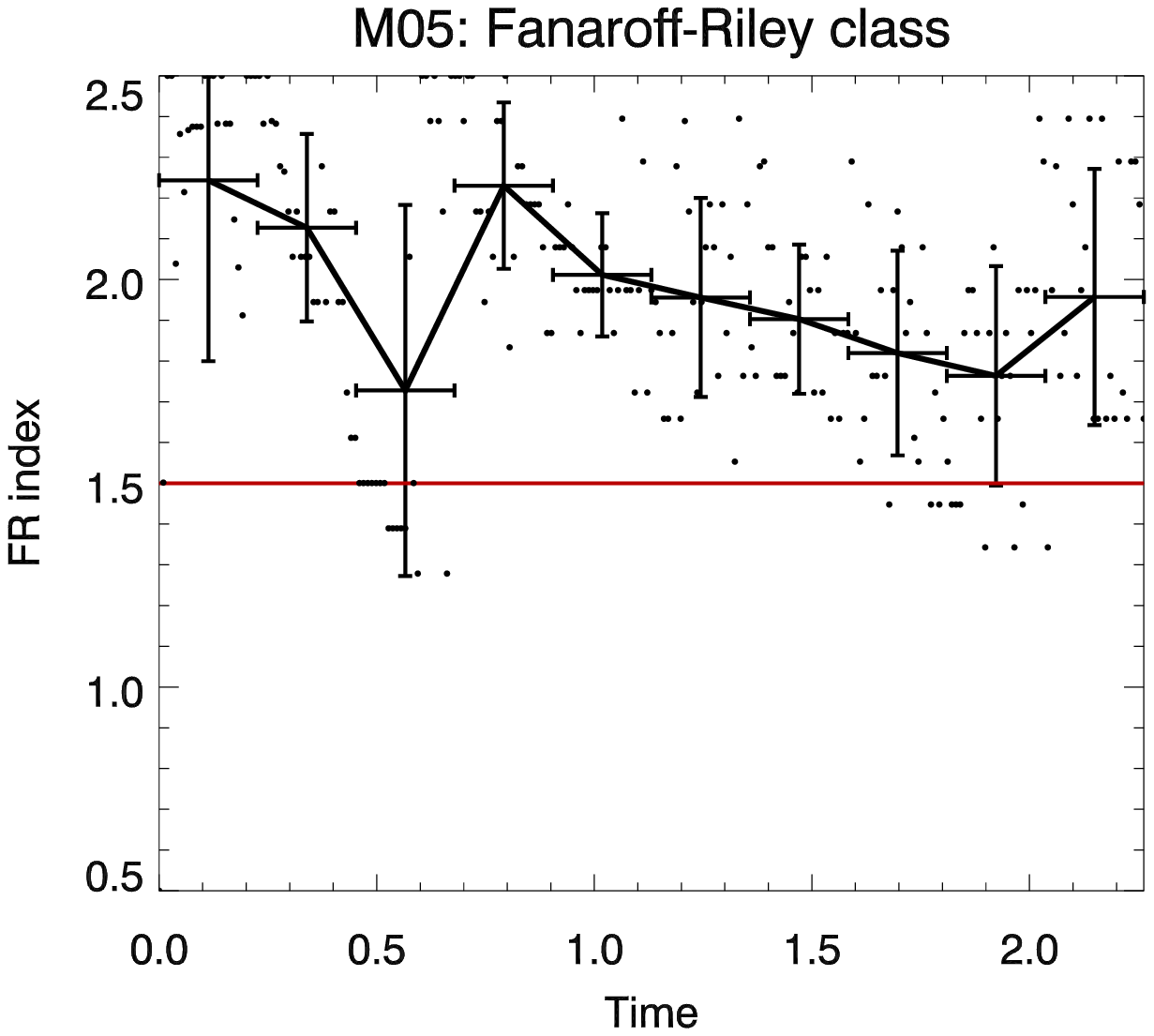} 
       \includegraphics[width=.48\textwidth]{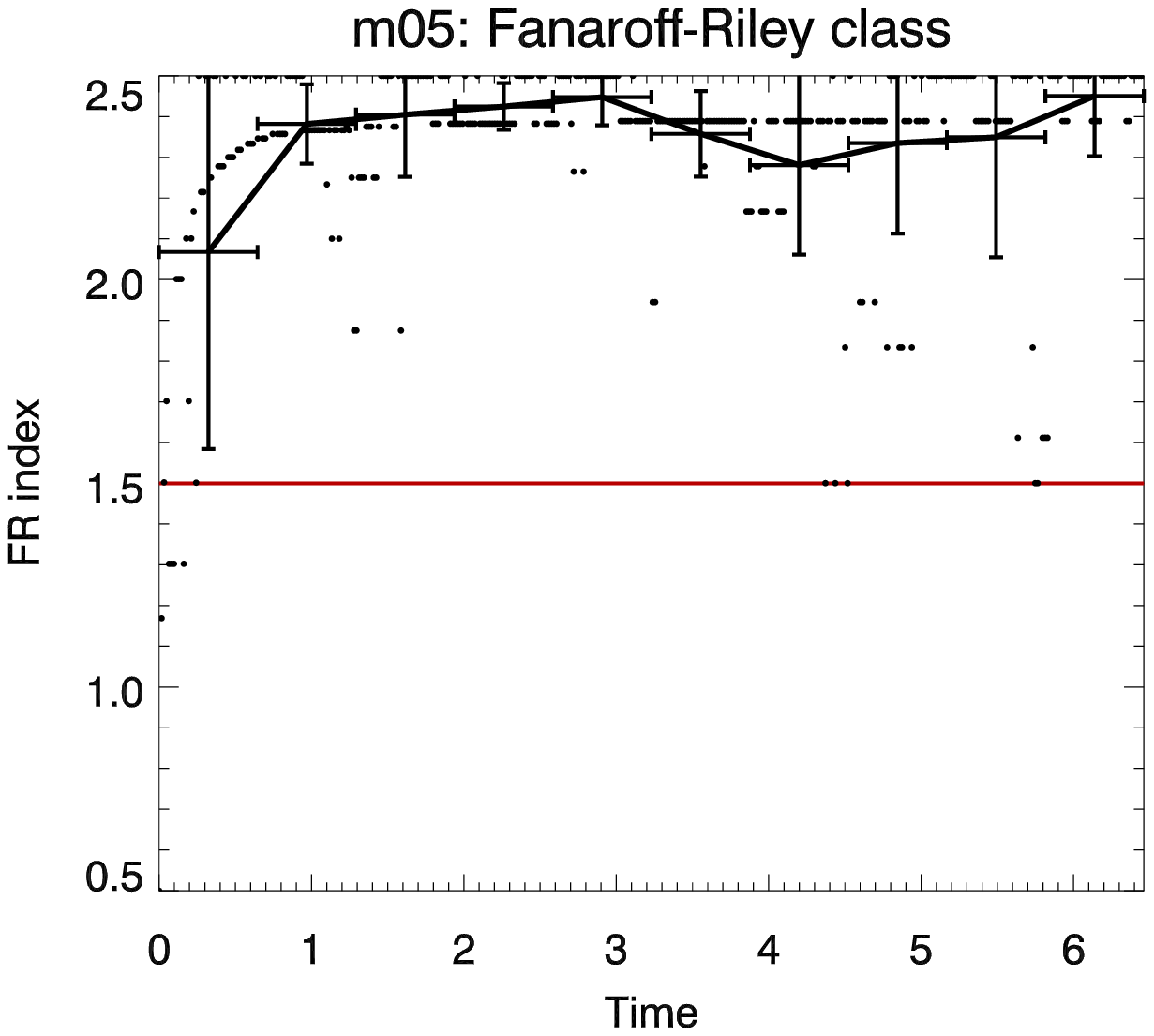} \\ 
       \includegraphics[width=.48\textwidth]{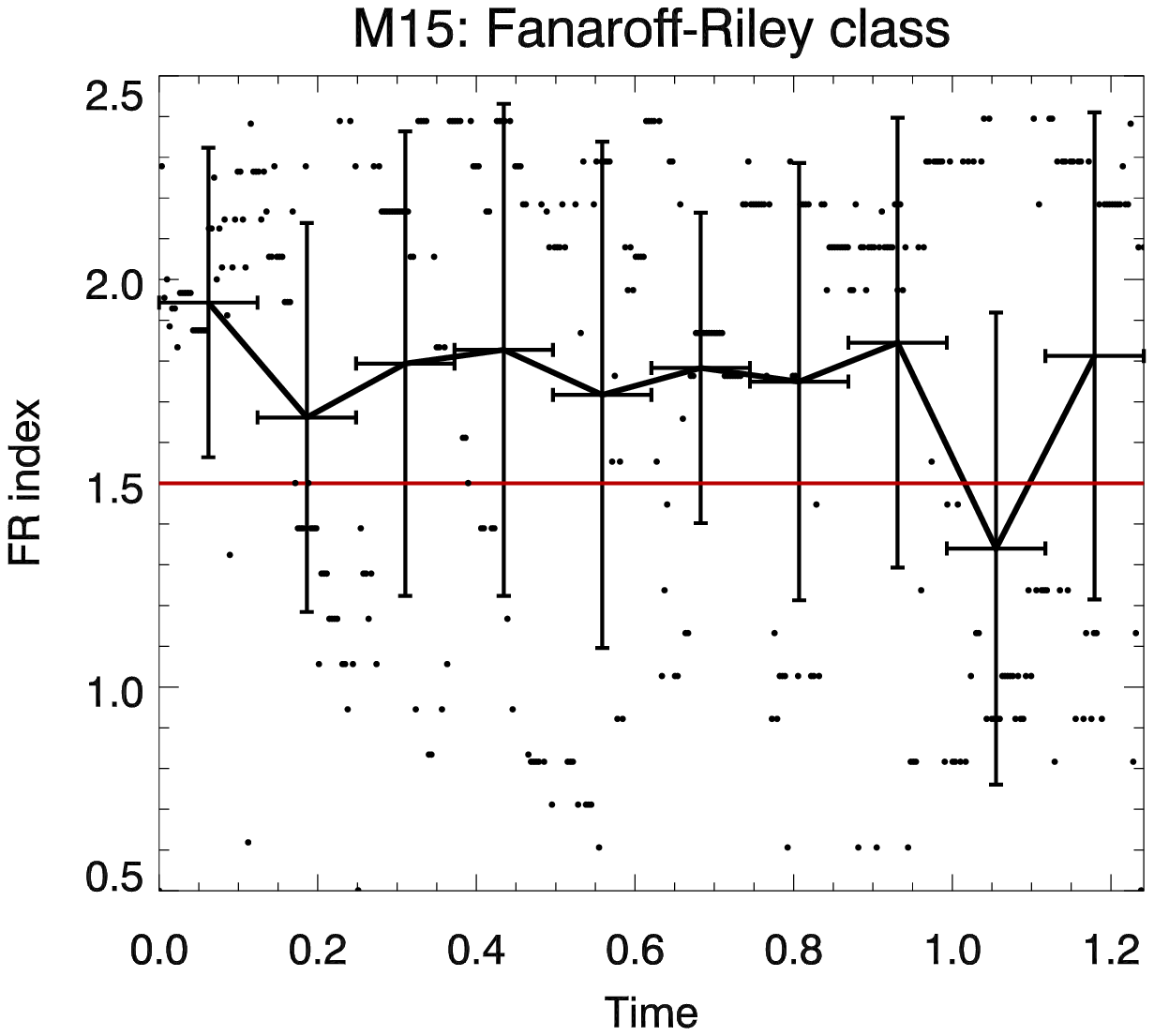} 
       \includegraphics[width=.48\textwidth]{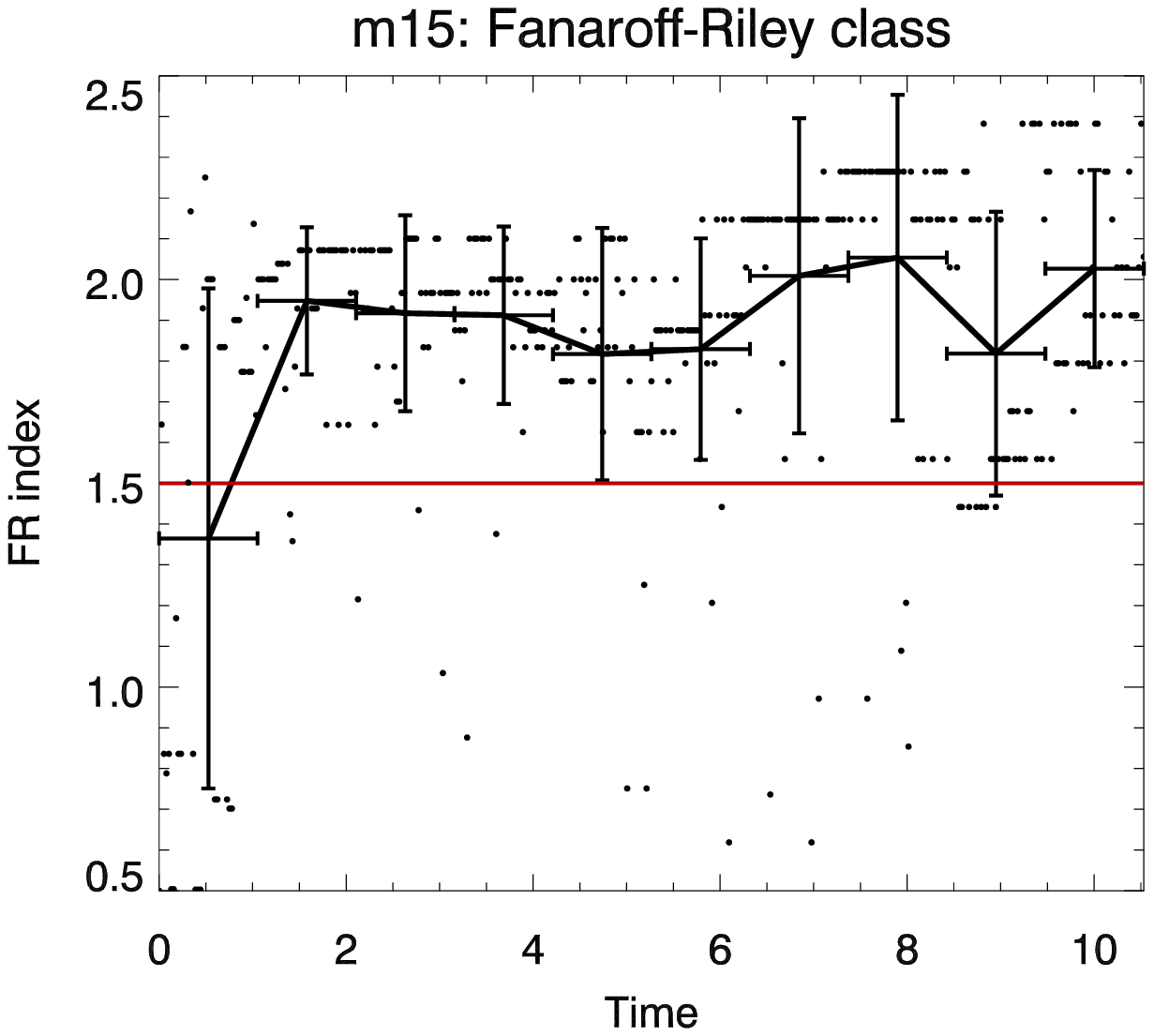} \\
       \includegraphics[width=.48\textwidth]{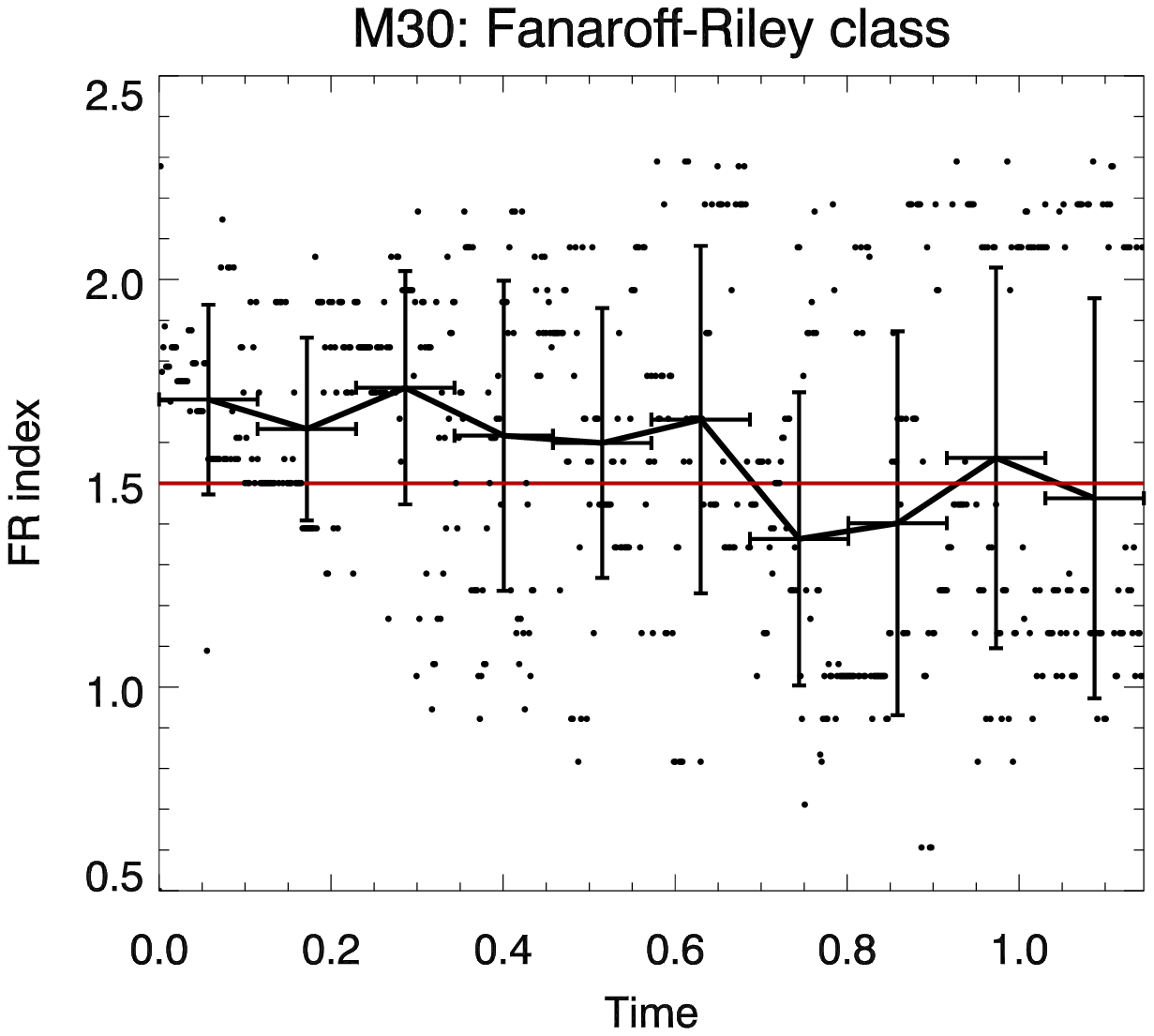} 
       \includegraphics[width=.48\textwidth]{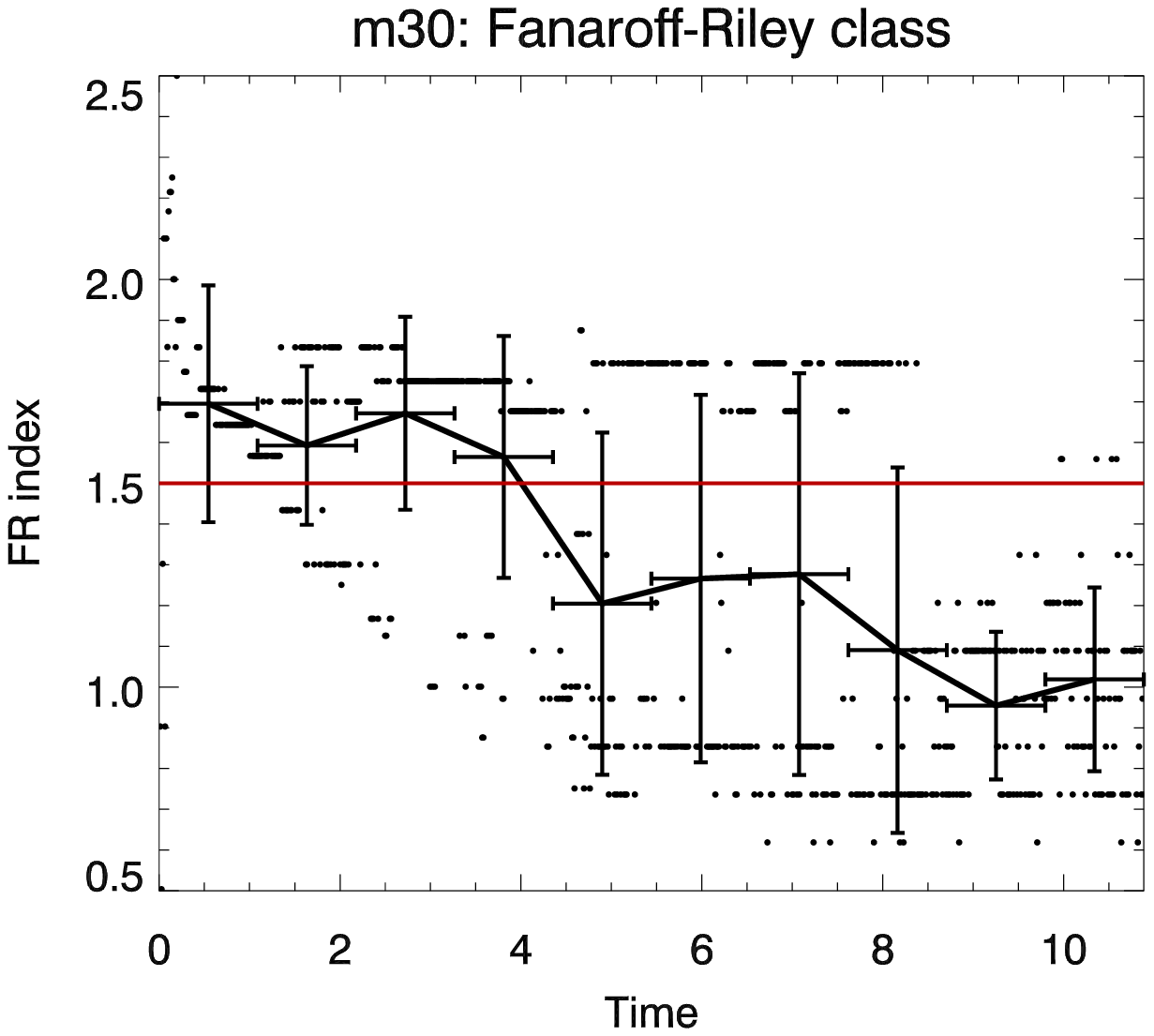} \\
     \vspace*{0pt}
 \caption{Fanaroff-Riley class over time for all simulations. The
   solid line with error bars denote the mean value and standard
   deviation for the given interval. Additionally, all snapshots are
   represented by filled circles.
   The red line denotes the
   border between FR~I and FR~II. See text for more details.}
\label{fig:frc}
\end{figure*}

We show FR against simulation time in Figure~\ref{fig:frc}. For runs~M05
and~M15, the average FR is almost always clearly above \change{1.7}, 
indicating an FR~II source,
as expected. The values below 1.5 \change{are} mostly due to shocks near
the inner boundary, which sometimes get quite strong, especially
because of the reflecting boundary conditions we use. This artifact of
the procedure does however not \change{destroy} the general trend.
We also find a clear FR~II morphology for 
the heavy low opening angle jets m05
and m15. Run~m30 is an FR~II source until it reaches its terminal
shock limit $L_\mathrm{1b}$ \change{around $t=4$}. From then onwards, it changes its
character to an FR~I radio source. For run~M30, the FR index is on average
above 1.5 for times earlier then~$t=0.7$, and below afterwards
\change{-- but only marginally so}. 
So, it makes principally the same transition as run~m30\change{, but
  the transition is much less marked. In this case the }
jet is quite underdense\change{,} form\change{ing} a prominent
cocoon, which protects it against entrainment of the ambient
medium\change{; it}  therefore transport\change{s} more energy to its head
region than the m30 \change{jet}. Consequently, the head region 
\change{also contains
strong shocks,} and the emission maximum is \change{often} located there.
We expect that th\change{e} FR~II branch of the diagram, which is already
weaker than the FR~I branch after $t=0.7$, would die out completely,
as in run m30,
if there were stronger entrainment. This could happen either when the
cocoon approaches the pressure of the environment, or if it had a
sufficient source of gas, for example \change{a gas disk in its
  equatorial plane.}

\section{Discussion}\label{sec:disc}

We have demonstrated both analytically and by hydrodynamics
simulation that for jets collimated by the external pressure,
the large-scale morphology develop\change{ed} as a result of the
interaction of the jet with its surroundings near its collimation
region.
 The ratios of three critical
length scales to one another control the behaviour. Only two of them
are independent, and they are set by the jet's opening angle and external
Mach-number. 

We find that light jets with cocoons form for high
external Mach-numbers and moderate to \jmr{large} opening angles. It is well
known that the majority of extragalactic \jmr{radio sources have} extended radio
lobes and \jmr{the jets} are therefore very light compared to the external medium. 

We predict that all lobed sources go through a
phase in which hot spots occur near the tip of the lobe. If the jets collimate 
before \change{their} ram-pressure falls
below the thermal pressure of the environment, then the jet retains
FR~II morphology. In our 
combined analytical / simulation analysis, this depends on the initial
opening
angle of the jet with a transition for $\theta=24^\circ$.
Our simulations with $5^\circ$ and $15^\circ$ confirm this scenario,
and consistently develop FR~II morphology.

The jets with
$30^\circ$ run out of thrust, before the re-collimation shock reaches
the axis. This result does not depend on the external Mach-number, which
differs by a factor of~100 between our \change{two 30$^\circ$} runs.

Run~m30 has high opening angle and low external Mach number. This
results in a collimated jet density that is only slightly below the
ambient density. Consequently, the cocoon is not as pronounced as in
the higher external Mach number runs, and is not able to
shield the jet from the ambient medium. Instabilities at the
re-collimation shock, which are pronounced at high opening angle,
entrain ambient gas directly in the jet in this run. This slows the
jet down, which produces \change{a} conical turbulent flow downstream of the
re-collimation shock. Because of the large opening angle, the ram
pressure of the jet is insufficient to propagate the terminal shock
through the re-collimation shock. In effect, the jet shows FR~II
morphology until the terminal shock reaches its limiting location, and
appears as \change{an} FR~I source afterwards.

Run~M30 has a very similar shock structure \change{to that of} run ~m30. We also
classify it on average as FR~I at late times. This
suggests that the transition from class~II to class~I is independent
of the jet velocity, which differs by two orders of magnitude between
the runs, but only depends on the initial opening angle.

The cocoon of run~M30 differs from th\change{at} of run~m30: it has lower
density and at the end of the simulation it still effectively prevents 
entrainment of ambient gas. As argued above, this also leads
frequently (but still in the minority of snapshots)
to dominant shocks in the head region. 
M30's cocoon could transform to a morphology similar to \change{that
  of} run~m30
if it entrain\change{ed} sufficient amounts of dense ambient gas\change{.} 
The cocoon itself is stabilised against entrainment of
ambient gas due to Rayleigh-Taylor instabilities by its deceleration
\citep{Alex02}. The deceleration up to a source size of $L_2$ can be
approximated by a power law \citep[e.g.][]{KA97}. This 
\change{power law} breaks down
when the source approaches 
$L_2$\change{, and the sideways expansion stalls} 
(\citeauthor{Gaiblea09} \citeyear{Gaiblea09} show this for FR~II jet
simulations). The cocoon
\change{is then Rayleigh-Taylor unstable and} will be refilled with
ambient gas, which will also lead to
entrainment into the main beam (i.e. the forward flow downstream of
the re-collimation shock). The cocoon will transform accordingly and
this should result in a source structure similar to the m30 run.
Our M30 run is however a factor of 100 away from this transition
point, and for the present study we did not have the computational
resources to simulate th\change{i}s far. 
\citet{KB07} invoke this \change{expansion to $L_2$ and the subsequent
entrainment} as the sole mechanism to
account for the FR dichotomy. 
 
Observations of the FR~I source Centaurus~A \citep[e.g.][]{Hardea07} suggest 
another source of entrainment\change{.} If the re-collimation shock is located
inside a dense gas disk, that gas disk would be difficult to push
aside, and the dense gas might come close to the \change{beam} 
\citep{Neumea07}, especially if
the latter precesses \citep{Morgea99}. Entrained gas can be resolved
in this source \citep[e.g.][]{TL09}. 
\change{This confirms that entrainment might be an important factor
  for a radio source to show FR~I morphology.} 
Centaurus~A shows evidence for a high Mach
number bow shock \citep{Crostea09}, and is therefore still far from pressure equilibrium
with its surroundings. 
\change{A mechanism as proposed by \citep{KB07}, which relies entirely
  on the radio source to be close to pressure equilibrium with its
  surroundings, therefore seems to have difficulties to explain the
  observations of Centaurus~A.}
Stellar winds within the jet \citep{Kom94} are another
possib\change{le mechanism by which} dense gas 
\change{is brought} to the jet plasma.

While entrainment has been recognised in many studies 
\change{as being} important
for the emergence of an FR~I morphology, we find there is another,
necessary condition \change{--} a large opening angle \change{for} the
jet. Regarding observed opening angles, probably the best studied
cases are M87 (FR~I) and Cygnus~A (FR~II). 
\citet{JBL99} present an observational study of the inner jet in M87\change{.}
The jet starts out at an
apparent half opening angle of about $30^\circ$ at about 0.04~pc
distance from the core. Its opening angle then reduces gradually to
about $7^\circ$ about 14~pc distance from the core. In our
simulations, collimation also happens gradually, over a
comparable length scale ratio of about two orders of magnitude.
\citet{Gracea09} present a
detailed analysis of the inner jet of M87, using a
magnetohydrodynamics-based fit to the observational data. The
available data constrains their models \change{well}. Among 2600 MHD models,
they find several that fit the data well, apart from the
innermost point with the largest opening angle. Discarding the
innermost \change{data point} yields a second innermost opening angle of only
about $17^\circ$. While this might first appear to contradict our
results (see below), the other features of their best\change{-}fitting models
are in good agreement\change{.} They find that the limb brightening of the jet
is due to enhanced intrinsic emission, rather than a beaming
effect due to velocity structure. Further, they identify the bright
feature 'knot~A' with a re-collimation shock. If the jet density is
related to the emissivity, limb brightening before the re-collimation
shock reaches the axis is predicted by our simulations. It is a
natural consequence of the transition from a conical to a pressure
collimated jet, which has already been noted by \citet{KF98}.

The opening angles derived from observations cannot be directly
compared to the initial opening angles we impose in our simulations or
use in our analytic work. This is because the jet base is both hard
to resolve and might not display bright features over the full jet
width. The first reflection point of the re-collimation shock on the
axis, or for FR~I sources the flaring point will be locations of
bright emission, where an opening angle could easily be measured. \change{Using
this point, we obtain} measured half opening angles of 
$\approx5^\circ$ for our simulations of $30^\circ$ initial opening
angle. The half opening angle at knot~A in M87 is $3^\circ$
\citep{JBL99}. \change{The agreement with our results is reasonably good for
M87, given that the observational value is a lower limit, since the
full width at quarter maximum is used, and in addition, our simulations do not take
into account magnetic and relativistic effects.}

Measuring the opening angle at the first reflection point of the
re-collimation shock yields a half opening angle of  \change{$\approx 1^\circ$}
for runs~M05 and ~M15. \citet{KF98} present similar simulations to our low
opening angle cases and compare \change{then} in detail to the FR~II radio source
Cygnus~A. From Cygnus~A's measured cocoon aspect ratio, they predict
an initial half opening angle between $11.5^\circ$ and
$15.25^\circ$. They identify knot~3 at 8~arcsec (about 6.4~kpc)
distance from the core as the first
reflection point of the re-collimation shock. As in the case of M87,
they mention the limb brightening to support this identification. They quote an opening
angle of $0.8^\circ$, measured at 20~arcsec from the core\change{, pointing} out
that the measurement is inconsistent with their predicted initial opening
angle. However, when doing the measurement in a similar way \change{for} the
simulation and \change{for} the observation, the opening angle on the scale of
the re-collimation shock is actually very similar.

The most important parameter that correlates with the observed jet morphology
is the jet power. We find that only
jets with a large initial opening angle have the potential to form
an FR~I source. This suggests an anti-correlation of the opening angle
with the jet power.
\citet{Pushea09} present a sample of parsec scale radio sources, all
having opening angles of order a few degree\change{s}. Interestingly,
they find an anti-correlation between the opening angle and the
Lorentz-factor, which is consistent with the jet acceleration
model of \citet{Komea07}.  
Therefore, there might indeed be a correlation of a jet's opening
angle with its power.
We are not aware of a systematic study of the correlation of the
jet opening angle with FR class.

We identify the quasi-stationary terminal shock in the
high opening angle runs with the flaring point in FR~I jets. 
This is in line with other theoretical work \citep[e.g.][]{Bick84,PM07}.
\change{Observationally,} \citet{Hardea05} and
\citet{Jetea05} \change{find} that the flaring point is located in a region of
flattening ambient pressure profile
\change{-- a result that can be understood in the sense that in such a region}
it is much more likely \change{that} the jet's
ram-pressure \change{will} become equal to the ambient pressure.

If the identification of the flaring point with $L_\mathrm{1c}$ \change{is}
correct, the location of the flaring point would constrain
$L_1$. Together with information on the jet speed, the external
density and pressure, this would provide \change{the} possibility
\change{of} measur\change{ing} the power of FR~I jets.

We do not expect that the general features we discuss here would
change if we \change{had} done the simulations in three dimensions.
\citet{KF98} have done simulations of low opening angle axisymmetric
FR~II jets for both the relativistic and the non-relativistic case,
with very comparable results. The critical angle for the morphological
transition depends on the angle that the re-collimation shock makes
with the axis. In this respect, our results for $5^\circ$ and
$15^\circ$ opening angle are quite similar to their $5^\circ$ and
$20^\circ$ cases. Also their relativistic simulations show 
quite similar behaviour. The critical angle of $24^\circ$ we derive is
consistent with the fact that \citet{KF98} still find FR~II structure at
$20^\circ$ for their relativistic and non-relativistic cases alike.
While the value of the critical angle might depend on where the
re-collimation shock exactly meets the axis, which may well depend on
details \change{and also} possibly on the magnetic field, we expect the general
result to be robust.

\section{Conclusions}\label{sec:conc}

We have shown that the \pa{large-scale} morphology of pressure
collimated jets is determined by the
processes near the region where the jet collimates due to the ambient
pressure. 
We find that the main requirement to form a very underdense jet, and
therefore a wide radio lobe is a high external Mach number.
We find a clearly distinct shock structure
for jets with initial half opening angles below and above 
\change{about} $24^\circ$, respectively. 
Jets with lower opening angles always have FR~II morphology, and we
classify them accordingly by an appropriate FR-index based on the
shock structure.
Higher opening angle jets are found to transition from class~II to
class~I. 
In the latter case, entrainment is identified as an important factor
\change{in} the jet morphology. The initial opening angle 
\change{near the jet base}
is much larger than the
observed one. Predicting the observed opening angle from our
simulations results in values which agree well with observations of
M87 and Cygnus~A.

\section*{Acknowledgements} The software
used in this work was developed by the DOE-supported
ASC/Alliance Center for Astrophysical Thermonuclear Flashes at the 
University of Chicago.\\
\bsp
\setlength{\labelwidth}{0pt}
\bibliographystyle{mn2e}
\bibliography{/Users/mkrause/texinput/references}

\label{lastpage}

\end{document}